\documentclass[11pt]{article}
\usepackage[dvips]{color}                 
\usepackage{graphicx}                     
\usepackage{amssymb}
\usepackage{dsfont}
\usepackage{amsmath}
\renewcommand{\today}{\ifcase\day\or 1st\or 2nd\or 3rd\or 4th\or 5th\or 6th\or
 7th\or 8th\or 9th\or 10th\or 11th\or 12th\or 13th\or 14th\or 15th\or 
 16th\or 17th\or 18th\or 19th\or 20th\or 21st\or 22nd\or 23rd\or 24th\or
 25th\or 26th\or 27th\or 28th\or 29th\or 30th\or 
 31st\fi~\ifcase\month\or January\or February\or March\or April\or
 May\or June\or July\or August\or September\or October\or November\or
 December\fi \space \number\year}   
\newcommand{\mytitle}[1]{
                         \begin{center}
                           \LARGE{\textbf{#1}}
                         \end{center}}
\newcommand{\myauthor}[1]{\textbf{#1}}
\newcommand{\myaddress}[1]{\textit{#1}}
\newcommand{\mypreprint}[1]{\begin{flushright} #1 \end{flushright}}
\begin{document}

\begin{titlepage}
\mypreprint{
TUM-T39-05-15 \\
}

\vspace*{0.5cm}
\mytitle{Signatures of chiral dynamics in the Nucleon to Delta transition\footnote{This research is part of the EU Integrated Infrastructure 
Initiative Hadron Physics under contract number RII3-CT-2004-506078. This work has also been supported by BMBF.}}
  \vspace*{0.3cm}

\begin{center}
 \myauthor{Tobias A. Gail} and
  \myauthor{Thomas R. Hemmert}, 

  \vspace*{0.5cm}
\myaddress{Theoretische Physik T39, Physik Department\\
	TU-M\"unchen, D-85747 Garching, Germany}
  \vspace*{0.2cm}
\end{center}

\vspace*{0.5cm}

\begin{abstract}
\noindent
Utilizing the methods of chiral effective field theory we present an analysis of the electromagnetic $N\Delta$-transition current in the 
framework of the non-relativistic ``small scale expansion'' (SSE) to leading-one-loop order. We discuss the momentum dependence of the 
magnetic dipole, electric quadrupole and coulomb quadrupole transition form factors up to a momentum transfer of $Q^2<0.3$ GeV$^2$.
Particular emphasis is put on the identification of the role of chiral dynamics in this transition. Our analysis indicates that there is indeed
non-trivial momentum dependence in the two quadrupole form factors at
small $Q^2<0.15$ GeV$^2$ arising from long distance pion physics,
leading for example to negative radii in the (real part of the) quadrupole transition form factors.
We compare our results with the EMR($Q^2$) and CMR($Q^2$) multipole-ratios from pion-electroproduction experiments and 
find a remarkable agreement up to four-momentum transfer of $Q^2\approx 0.3$ GeV$^2$. Finally, we discuss the chiral extrapolation of the
three transition form factors at $Q^2=0$, identifying rapid changes in the (real part of the) quark-mass dependence of the quadrupole 
transition moments  for pion masses below 200 MeV, which arise again
from long distance pion dynamics.  Our findings indicate that dipole
extrapolation methods currently used in lattice QCD analyses of
baryon form factors are not applicable for the chiral extrapolation of
$N\Delta$ quadrupole transition form factors.
\end{abstract}
\end{titlepage}

\vfill\pagebreak
\section{Introduction}
\setcounter{footnote}{0}

$\Delta$(1232) is the lowest lying baryon resonance with quantum numbers spin S=3/2 and isospin I=3/2. It can be studied, for example, in
 the process of pion-photoproduction on a nucleon. Therein it shows up both as a clear signal in the cross section and as a pole at
 $M_\Delta=1210-i\,50$ MeV \cite{PDG} in the complex W-plane, with W=$\sqrt{s}$ denoting the total energy
 as a function of the Mandelstam variable $s$. At the position of the resonance the incoming photon can excite the target nucleon
into a $\Delta$(1232) resonant state via a M1 or an E2 electromagnetic
multipole transition. Assuming Delta-pole dominance at this
 energy, one can relate the pion-photoproduction multipoles describing the final $\pi N$-state of this process to the strengths of the
 sought after $\gamma N\Delta$-transition moments. Extensive research over the past decades has produced the result   
EMR$=-(2.5\pm0.1_{\textnormal{stat}}\pm0.2_{\textnormal{sys}})\%$ \cite{Beck}, demonstrating that in this ratio of quadrupole to dipole transition strength the 
magnetic dipole dominates the transition to the percent level. Extending these studies to pion-electroproduction the incoming (virtual) photon
carries a four-momentum (squared) $q^2<0$ and 
can also utilize a charge-quadrupole (C2) transition to produce an intermediate $\Delta$(1232) resonance. The three electromagnetic
multipole transitions M1, E2 and C2 then become functions of momentum transfer (squared) $q^2$, analogous to the well-known
 electromagnetic form factors of the nucleon studied in elastic
electron scattering off a nucleon target. Extensive experimental
studies of pion-electroproduction in the $\Delta(1232)$ resonance region
\cite{Stein,Baetzner,Bartel,Bonn} have already
demonstrated that also for finite $q^2$ both the electric and
the coulomb $N\Delta$ quadrupole transitions remain
 ``small'' 
(at the percent level) compared to the dominant magnetic $N\Delta$ dipole transition. However, recently, new high precision studies at
 continuous beam electron machines \cite{CLAS,OOPS,MAMI}
 have been performed in order to quantify the observed dependence of these transitions with respect to $q^2$. It is hoped that from these new
 experimental results one can infer those relevant degrees of freedom within a nucleon which are responsible for the observed (small) quadrupole
 components in the $\gamma N\Delta$-transition.
 The theoretical study presented here attempts to identify the active degrees of freedom in these three transition form factors
 in the momentum region $Q^2=-q^2<0.3$ GeV$^2$.  

Historically the non-zero strength of the $N\Delta$ quadrupole transitions has raised a lot of interest because such transitions are absent in
(simple) models for nucleon wave functions with spherical symmetry. The issue of detecting a "deformed shape" of the nucleon
via well-defined observables in scattering experiments, however, is intriguing to the minds of nuclear physicists up to this day, {\it e.g.}
 see the discussion in ref.\cite{bernstein}. On the theoretical side, most of the work over the past 20 years has focused on the idea that a
 ``natural'' explanation for the non-zero 
$N\Delta$ quadrupole transition moments could arise from pion-degrees of freedom present in the nucleon wave function, {\it i.e.} from the
 so-called 
``pion-cloud'' around the nucleon. Many calculations to quantify this hypothesis have been pursued, within the 
skyrme model ansatz ({\it e.g.} see \cite{WW}), within dynamical pion-nucleon models ({\it e.g} see \cite{SL}), within quark-meson coupling
models ({\it e.g.} see \cite{Alfons}), within chiral bag models ({\it e.g} see \cite{Lu}), within chiral quark soliton models ({\it e.g.} see \cite{Sol}),
... to name just a few of them.  Around 1990---based on the works of refs.\cite{GSS,JM,BKKM}---the qualitative concept of the ``pion cloud''
 around a nucleon could be put on a firm field-theoretical footing within
the framework of chiral effective field theory (ChEFT) for baryons. The pioneering study of the strength of the electric $N\Delta$ quadrupole
transition within ChEFT for real photons was performed in  ref.\cite{BSS}, and the first calculation of all three $N\Delta$-transition form factors
for $Q^2<0.2$ GeV$^2$ within the SSE-scheme of ChEFT \cite{HHK} was given in ref.\cite{GHKP}. In this paper we present an
 update and extension of the non-relativistic ${\cal O}(\epsilon^3)$ SSE calculation of
the latter reference and compare the results both to experiment as well
as  to recent theoretical calculations \cite{DMTMAID}.

Before we begin the discussion of the general $\gamma N\Delta$ transition matrix element in the next section, we want to remind the reader,
 that the ``pion cloud'' around the nucleon in ChEFT calculations does not just lead to non-zero quadrupole transition form factors, but is also
 responsible for the fact that {\em all three $N\Delta$-transition form factors}---unlike the case of the elastic nucleon form factors, see {\it e.g.} 
ref.\cite{BFHM}---are {\em complex valued}, due to the presence of the open $\pi N$-channel, in accordance with ref.\cite{JS}. In the following 
we continue this paper
with a brief discussion of the effective field theory calculation in
section \ref{sec:calc} and present our results in section \ref{sec:res} before summarizing our main
 findings in section \ref{sec:conc}. A few technical aspects are relegated to two appendices.

\section{Parametrization of the matrix element}
\label{sec:param} 

Demanding Lorentz covariance, gauge invariance and parity conservation
the matrix element of a $I\left(J^P\right)=\frac{3}{2}\left(\frac{3}{2}^+\right)$ to 
$\frac{1}{2}\left(\frac{1}{2}^+\right)$ transition can be parametrized
in terms of three form factors. For our calculation we follow the conventions of ref.\cite{GHKP} and choose the definition:
\begin{eqnarray}
i\mathcal{M}_{\Delta\rightarrow N\gamma} & = &+\sqrt{\frac{2}{3}}
\,\frac{e}{2M_N}\bar{u}(p_N)\gamma_5\bigg[G_1(q^2)
(\not\!q\epsilon_{\mu}-\not\!\epsilon
q_{\mu})+\frac{G_2(q^2)}{2M_N}(p_N\cdot\epsilon
q_{\mu}-p_N\cdot q\epsilon_{\mu}) \nonumber
\\&&+\frac{G_3(q^2)}{2\Delta}(q\cdot
\epsilon q_{\mu}-q^2\epsilon_{\mu})\bigg]u^{\mu}_{\Delta}(p_{\Delta}).
\label{defff}
\end{eqnarray}
Here $e$ denotes the charge of the electron and $M_N$ is the mass of a nucleon, $p_{N/\Delta}^{\mu}$ denotes the
relativistic four-momentum of the outgoing nucleon/incoming $\Delta$ and $q_{\mu}$,
$\epsilon^{\mu}$ are the momentum and polarization vectors of the outgoing photon,
respectively. As discussed in ref.\cite{GHKP} the small scale $\Delta=M_\Delta-M_N$ denoting the nucleon-Delta mass-splitting had to 
be introduced in front of the $G_3(q^2)$ form factor in order to obtain a consistent matching between the calculated 
$\Delta\rightarrow N\gamma$ 
amplitudes and the associated $N\Delta$-transition current at
leading-one-loop order in the ChEFT framework of SSE \cite{HHK}. The dynamics of the outgoing nucleon is 
described via a Dirac spinor $u(p_N)$, while the associated $\Delta$(1232) dynamics is parametrized via a Rarita-Schwinger spinor 
$u_\mu(p_\Delta)$. From the point of view of chiral effective field theory the
signatures of chiral dynamics in the $N\Delta$-transition are particularly transparent in the $G_i(q^2),\,i=1,2,3$ basis, which serves as the 
analogue of the
Dirac- and Pauli-form factor basis in the  vector current of a nucleon. However, most experiments and most model calculations refer to the
multipole basis of the general $N\Delta$-transition current.
The allowed magnetic dipole, as well as electric- and charge quadrupole
transitions are parametrized via the form factors $\mathcal{G}_M^*(q^2)$,
$\mathcal{G}_E^*(q^2)$ and $\mathcal{G}_C^*(q^2)$ defined by Jones and
Scadron \cite{JS}. They are connected to our choice via the relations
\begin{eqnarray}
\mathcal{G}_M^*(q^2) & =
	&\frac{M_N}{3(M_N+M_{\Delta})}\bigg[\left((3M_{\Delta}+M_N)(M_{\Delta}+M_N)-q^2\right)
	\frac{G_1^{\dagger}(q^2)}{2M_NM_{\Delta}}-(M_{\Delta}^2-M_N^2-q^2)
	\frac{G_2^{\dagger}(q^2)}{4M_N^2}\nonumber \\&&-q^2\frac{G_3^{\dagger}(q^2)}{2M_N\Delta}\bigg],\label{con1} \\
\mathcal{G}_E^*(q^2) & = &
	\frac{M_N}{3(M_N+M_{\Delta})}\bigg[(M_{\Delta}^2-M_N^2+q^2)
	\frac{G_1^{\dagger}(q^2)}{2M_NM_{\Delta}}-(M_{\Delta}^2-M_N^2-q^2)\frac{G_2^{\dagger}(q^2)}{4M_N^2}-
	q^2\frac{G_3^{\dagger}(q^2)}{2M_N\Delta}\bigg], \label{con2}\\
\mathcal{G}_C^*(q^2) & =
&\frac{2M_N}{3(M_N+M_{\Delta})}\bigg[\frac{M_{\Delta}}{M_N}G_1^{\dagger}(q^2)
	-(M_{\Delta}^2+M_N^2-q^2)\frac{G_2^{\dagger}(q^2)}{4M_N^2}-
	(M_{\Delta}^2-M_N^2+q^2)\frac{G_3^{\dagger}(q^2)}{4M_N\Delta}\bigg]\label{con3}. 
\end{eqnarray}
As these multipole form factors have been defined for the
$N\gamma\rightarrow\Delta$ reaction they are linear combinations of
the hermitian conjugate $G_i^{\dagger}(q^2)$ form factors.\\
For a comparison with experimental results we also note that the notation of Ash \cite{Ash} is connected to the Jones-Scadron 
form factors via:
\begin{eqnarray}
\mathcal{G}_M^{*Ash}(q^2) & = & \frac{1}{\sqrt{1-\frac{q^2}{(M_N+M_{\Delta})^2}}}\mathcal{G}_M^{*JS}(q^2) \label{eq:ash}
\end{eqnarray}
The full information about the rich structure of the general (isovector) $N\Delta$-transition current is hidden in these three {\it complex} form 
factors. In experiment this transition is studied in the process $e\,p\rightarrow e^\prime\,N\pi$ in the region of the $\Delta$-resonance 
({\it e.g.} see ref.\cite{OOPS} and references given therein), which has access to a lot more hadron structure properties than just the 
$N\Delta$-transition current of Eq.(\ref{defff}). Based on 
the observation that the $\gamma^* N\rightarrow\Delta N\pi$ transition is dominated by the magnetic dipole transition and under the assumption that 
intermediate states are dominated by the (imaginary part) of the $\Delta$-propagator, one can relate three of the extracted (complex) 
pion-electroproduction 
multipoles in the isospin 3/2 channel $M_{1+}^{I=3/2}(W_{res},q^2),\,E_{1+}^{I=3/2}(W_{res},q^2),\,S_{1+}^{I=3/2}(W_{res},q^2)$ at the 
position of the resonance $W_{res}$ to the sought after form factors via 
\begin{eqnarray}
\textnormal{EMR} & \equiv &Re\left[\frac{E_{1+}^{I=3/2}(W_{res},q^2)}{M_{1+}^{I=3/2}(W_{res},q^2)}\right] \approx
	-\textnormal{Re}\left[\frac{\mathcal{G}_E^*(q^2)}{\mathcal{G}_M^*(q^2)}\right], \label{EMR} \\
\textnormal{CMR} & \equiv &Re\left[\frac{S_{1+}^{I=3/2}(W_{res},q^2)}{M_{1+}^{I=3/2}(W_{res},q^2)}\right] \approx
	-\frac{\sqrt{((M_{\Delta}+M_N)^2-q^2)((M_{\Delta}-M_N)^2-q^2)}}{4M_{\Delta}^2}
	\textnormal{Re}\left[\frac{\mathcal{G}_C^*(q^2)}{\mathcal{G}_M^*(q^2)}\right].\label{CMR}
\end{eqnarray}
Ultimately the validity of this (approximate) connection between the pion-electroproduction multipoles and the 
$N\Delta$-transition form factors has to be
checked in a full theoretical calculation. At present only nucleon- and $\Delta$-pole graphs have been included as intermediate states 
in calculations of the process 
$e\,p\rightarrow e^\prime\,N\pi$ in the $\Delta$ resonance region within chiral effective field theory ({\it e.g.} see ref.\cite{PV}). It remains to be seen to what extent non-resonant 
intermediate $N\pi$- or $\Delta\pi$-states in the isospin 3/2 channel of this process might lead to a correction\footnote{Such contributions
arise in a ${\cal O}(\epsilon^3)$ SSE calculation of the process $e\,p\rightarrow e^\prime N\pi$ in the $\Delta$(1232) resonance region 
\cite{GH}.}  in the connection between 
EMR, CMR and the form factor ratios as given in Eqs.(\ref{EMR},\ref{CMR}). As we cannot exclude this possibility at present, we have inserted
$\approx$-symbols in Eqs.(\ref{EMR},\ref{CMR}).
 
In the next step we will calculate the form factors of
Eq.(\ref{defff}) using chiral effective field theory and then discuss our results together with
experimental data for $\left|\mathcal{G}_M^{*Ash}(q^2)\right|$, EMR$(q^2)$
and CMR$(q^2)$. 

\section{Effective field theory calculation}
\label{sec:calc} 

The isovector $N\Delta$-transition current has been calculated to
${\cal O}(\epsilon^3)$ in non-relativistic SSE in ref.\cite{GHKP}. 
Here we briefly review the ingredients of this calculation.
The basic Lagrangean needed for a leading-one-loop calculation can be written as a sum of terms with increasing chiral
dimension. Divided into parts with different active degrees of
freedom it reads:
\begin{eqnarray}
\mathcal{L}_{SSE} & = & \mathcal{L}_{\pi\pi}^{(2)}+ \mathcal{L}_{\pi N}^{(1)}
+\mathcal{L}_{\pi\Delta}^{(1)}+\mathcal{L}_{\pi
N\Delta}^{(1)}+\mathcal{L}_{\gamma N\Delta}^{(2)}+\mathcal{L}_{\gamma
N\Delta}^{(3)}+...
\end{eqnarray}
In order to introduce a hierarchy of terms we utilize a counting
scheme called "small scale expansion" (SSE) \cite{HHK}. This scheme is based on
a triple expansion of the Lagrangean in the momentum
transfer $q^\mu$, the pion mass $m_{\pi}$ and the
Delta-nucleon mass-splitting in the chiral limit
$\Delta_0=M_{\Delta}-M_0$. It assigns the same
chiral dimension $\epsilon\in\{|q|,m_{\pi},\Delta_0\}$ to all three
small parameters.\\
The lowest order chiral Lagrangeans are given by \cite{HHK}:
\begin{eqnarray}
\mathcal{L}_{\pi\pi}^{(2)} & = &
	\frac{1}{4}F_{\pi}^2\left\{\textnormal{Tr}\left[\nabla_{\mu}U^{\dagger}\nabla^{\mu}U+
	\chi^{\dagger}+\chi U^{\dagger}\right]\right\}\label{Lag1},\\
\mathcal{L}_{\pi N}^{(1)} & = &\bar{N}\left[iv\cdot D+g_AS\cdot u\right]N\label{Lag2} ,\\
\mathcal{L}_{\pi\Delta}^{(1)} & = &
-\bar{T}_i^{\mu}\left[iv\cdot D^{ij}-\delta^{ij}\Delta_0+g_1S\cdot u^{ij}\right]g_{\mu\nu}T_j^{\nu}\label{Lag3},\\
\mathcal{L}_{\pi N\Delta}^{(1)} & = &
	c_A\left\{\bar{T}_i^{\mu}g_{\mu\alpha}u_i^{\alpha}N+\bar{N}{u_i^{\alpha}}^{\dagger}
	g_{\alpha\mu}T_i^{\mu}\right\}\label{Lag4}.
\end{eqnarray}
$g_A$, $g_1$
and $c_A$ denote axial nucleon, $\Delta$ and $N\Delta$ coupling constants in
the chiral limit, respectively, whereas $F_\pi$ corresponds to the pion decay constant.
The numerical values of these constants used throughout this work are
listed in table \ref{tab:param}. We are working in a non-relativistic 
framework
utilizing non-relativistic nucleon fields $N$ as well as non-relativistic Rarita-Schwinger fields $T_\mu^{i}$ for the four $\Delta$ states with 
isospin-indices $i,j$ \cite{HHK}. The pseudo Goldstone boson pion triplet $\pi^a$ is collected in the
$SU(2)$ matrix-valued field $U(x)$. The associated covariant derivatives for the pions $\nabla_\mu$, for the nucleon $D_\mu$ and for the
Deltas $D_\mu^{ij}$ as well as the chiral field tensors $\chi,\, u_\mu,\,u_\mu^{ij},\,u_\mu^{i}$ are standard and can be found in the literature 
\cite{HHK}. Finally, we note that $v_\mu$ denotes the velocity four
vector of the non-relativistic baryon and $S_\mu$ is the Pauli-Lubanski 
spin-vector of heavy baryon ChPT \cite{BKM}.

The local operators contributing to the $\gamma N\Delta$-transition up to order
$\epsilon^3$ are given in terms of the low energy constants $b_1$, $b_6$, $E_1$
and $D_1$ (see refs.\cite{HHK},\cite{GHKP}):
\begin{eqnarray}
\mathcal{L}_{\gamma N\Delta}^{(2)} & = &\frac{ib_1}{2M_0}
\bar{T}_i^{\mu}f_{\mu\nu}^{i+}S^{\nu}N+h.c. \label{Lag5} ,\\
\mathcal{L}_{\gamma N\Delta}^{(3)} & = &
	\frac{1}{4M_0^2}\bar{N}\Bigg[D_1g_{\mu}^{\nu}v^{\alpha}S^{\beta}f_{\nu\beta\alpha}^{i+}
	+2i\Delta_0 E_1f_{\mu\nu}^{i+}S^{\nu}+(b_1+2b_6)(S\cdot
	\overleftarrow{D})v^{\alpha}f_{\mu\alpha}^{i+}+\nonumber\\&&
	+(b_1-2b_6)v^{\alpha}f_{\alpha\mu}^{k+}\xi_{\frac{3}{2}}^{kj}S\cdot \overrightarrow{D}^{ji}+
	2b_1f_{\alpha\beta}^{k+}S^{\alpha}v^{\beta}\xi_{\frac{3}{2}}^{kj}\overrightarrow{D}_{\mu}^{ji}\Bigg]T_i^{\mu}+h.c.\label{Lag6}.
\end{eqnarray}
with $f_{\nu\beta\alpha}^{i+}=\frac{1}{2}Tr\left(\tau^{i}\left[D_\nu,f_{\beta\alpha}^+\right]\right)$
and $f_{\mu\nu}^+$ denoting the chiral field strength tensor of the external isovector background field \cite{BKM}.
The rich counter term structure contributing in this transition gives already an indication that the relevant scales governing the physics of 
the $N\Delta$-transition form factors 
arise from an interplay between long- and short-distance effects, making the detection of genuine signatures of chiral dynamics non-trivial 
in this transition.

Figure \ref{fig:feyn} shows all Feynman diagrams
contributing at ${\cal O}(\epsilon^3)$ in SSE. The strength of the contact terms in Fig.\ref{fig:feyn}a
is given by the LECs of Eqs.(\ref{Lag5},\ref{Lag6}), the
vertices and propagators appearing in the loop diagrams are determined
by the Lagrangeans Eqs.(\ref{Lag1})-(\ref{Lag4}). For details we are again referring to ref.\cite{GHKP}. 
Given that we are working in a non-relativistic formulation of SSE, the crucial step is the correct mapping of the $\Delta\rightarrow N\gamma$ 
transition amplitudes calculated from the diagrams of Fig.\ref{fig:feyn} to the form factors defined in Eq.(\ref{defff}). To ${\cal O}(\epsilon^3)$ 
one obtains (in the rest frame of the $\Delta$) \cite{GHKP}:
\begin{eqnarray}
i\mathcal{M}_{\Delta\rightarrow\gamma N} & = &
	\sqrt{\frac{2}{3}}\,e\,\bar{u}_v(r_N)\left[S\cdot\epsilon\,
	q_{\mu}\frac{G_1(q^2)}{M_N}+S\cdot
	q\,\epsilon_{\mu}\left(-\frac{G_1(q^2)}{M_N}-\frac{\Delta_0
	G_1(0)}{2M_N^2}\right.\right.\nonumber\\&&\left.\left.
	+\frac{\Delta_0G_2(q^2)}{4M_N^2}
	+\frac{q^2}{4M_N^2\Delta_0}G_3(q^2)\right)+S\cdot q\,
	v\cdot\epsilon \,q_{\mu}\left(\frac{G_1(0)}{2M_N^2}-\frac{G_2(q^2)}{4M_N^2}\right)\right.\nonumber
	\\&&\left.-S\cdot q\, \epsilon\cdot q\,
	q_{\mu}\frac{G_3(q^2)}{4M_N^2\Delta_0}+\mathcal{O}\left(\epsilon^4\right)
	\right]u_{v,\Delta}^{\mu}(0). \label{matching}
\end{eqnarray}
Eq.(\ref{matching}) provides the central connection between the diagrams and the sought after form factors. 
Calculating the ${\cal O}(\epsilon^3)$ diagrams of Fig.\ref{fig:feyn}
in non-relativistic SSE utilizing dimensional regularization one
obtains the
expressions given in appendix \ref{app:int}. The finite
parts in four dimensions (with renormalization scale $\lambda$) still
containing a Feynman parameter $x$ read: 
\begin{figure}
\begin{center}
\includegraphics[width=12cm,clip]{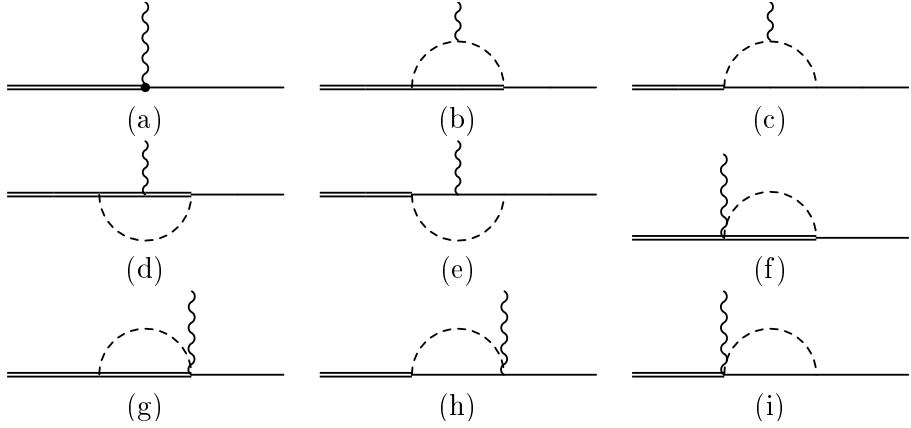}
\caption{The diagrams contributing to the $\Delta\rightarrow N\gamma$ transition at leading-one-loop
order in the SSE formalism \cite{GHKP}.}
\label{fig:feyn}
\end{center}
\end{figure}
\begin{eqnarray}
G_1(q^2) & = & A(\lambda)+\frac{c_A M_N}{(4\pi
	F_{\pi})^2}\Bigg[\Delta_0\left(\frac{g_A}{3}-\frac{145g_1}{81}\right)-\frac{4}{9}\Delta_0
	\int_0^1\!dx\,\left(5g_1(x-3)-
	9g_A(x-1)\right)x\ln{\left(\frac{\tilde{m}}{\lambda}\right)}\nonumber\\&&
	+\frac{4}{9}\int_0^1\!dx\,\left(5g_1(x-3)I(-x\Delta_0,\tilde{m})+9g_A(x-1)I(x\Delta_0,\tilde{m})\right)\Bigg]
	+{\cal O}(\epsilon^4) \label{res1},\\
G_2(q^2) & = & B(\lambda)-C_sq^2+\frac{c_AM_N^2}{(4\pi
	F_{\pi})^2}\Bigg[\frac{4}{81}\left(27g_A-35g_1\right)+\frac{16}{9}(5g_1-9g_A)\int_0^1\!dx\,x(x-1)
	\ln{\left(\frac{\tilde{m}}{\lambda}\right)}\nonumber\\&&
	+\frac{16}{9}\int_0^1\!dx\,\frac{x^2(x-1)}{\tilde{m}^2-x^2\Delta_0^2}\Delta_0\left(
	5g_1I(-x\Delta_0,\tilde{m})+9g_AI(x\Delta_0,\tilde{m})\right)\Bigg]+{\cal O}(\epsilon^4) ,\label{res2}\\
G_3(q^2) & = & \frac{8}{9}\frac{c_AM_N^2\Delta_0}{(4\pi
	F_{\pi})^2}\int_0^1\!dx\,\frac{x(2x^2-3x+1)}{\Delta_0^2x^2-\tilde{m}^2}\left(5g_1I(-x\Delta_0,\tilde{m})
	+9g_AI(x \Delta_0,\tilde{m})\right)+{\cal O}(\epsilon^4) \label{res3}.
\end{eqnarray}
The corresponding ${\cal O}(\epsilon^3)$ results in (non-relativistic)
SSE for the $N\Delta$-multipole transition form factors $\mathcal{G}_M^*(q^2)$,
$\mathcal{G}_E^*(q^2)$, $\mathcal{G}_C^*(q^2)$ are obtained by inserting 
Eqs.(\ref{res1}-\ref{res3}) into the definitions Eqs.(\ref{con1}-\ref{con3}).  
In Eqs.(\ref{res1}-\ref{res3}) we have introduced the quantities
$\tilde{m}^2 = m_{\pi}^2-q^2x(1-x)$ \cite{BFHM} and
\begin{eqnarray}
I(\omega,m_{\pi}) & = & \left\{\begin{array}{l c l}
	\sqrt{\omega^2-m_{\pi}^2}\left(\ln{\left(\frac{\omega}{m_{\pi}}+\sqrt{\frac{\omega^2}{m_{\pi}^2}-1}\right)}
	-i\pi\right)&&
	\frac{\omega}{m_{\pi}}>1\\
	\sqrt{m_{\pi}^2-\omega^2}\arccos{\left(-\frac{\omega}{m_{\pi}}\right)} &\,\textnormal{for}\, &
	-1\le\frac{\omega}{m_{\pi}}\le 1 \\
	-\sqrt{\omega^2-m_{\pi}^2}\ln{\left(-\frac{\omega}{m_{\pi}}+\sqrt{\frac{\omega^2}{m_{\pi}^2}-1}\right)}&&
	\frac{\omega}{m_{\pi}}<-1.\end{array}\right.
\end{eqnarray}
$A(\lambda)$ and $B(\lambda)$ collect all short range physics contributing to
$G_1(0)$ and $G_2(0)$:
\begin{eqnarray}
A(\lambda) &= & -\frac{1}{2}b_1+(2E_1^{(r)}(\lambda)-D_1^{(r)}(\lambda))\frac{\Delta_0}{4M_N} ,\label{Al} \\
B(\lambda) & = &2b_6-D_1^{(r)}(\lambda). \label{Bl}
\end{eqnarray}
 The renormalization scale dependence of the counter terms cancels the one
from the loop contributions.
We note that the ${\cal O}(\epsilon^3)$ structure with the coupling constant $E_1$ in Eq.(\ref{Lag6}) arises naturally in SSE. Its infinite 
part is required for a complete 
renormalization of the ${\cal O}(\epsilon^3)$ result, whereas its
(scale-dependent) finite part $E_1^{(r)}(\lambda)$ cannot be observed independently from the 
coupling $b_1$. For nucleon observables the finite parts of terms like $E_1^{(r)}(\lambda)$ are required in order to guarantee the decoupling 
of the 
Delta-resonance in the limit $m_\pi/\Delta_0\rightarrow 0$ (see for
example the discussion given in ref.\cite{HW}). In $N\Delta$-transition quantities like $G_1(q^2)$ of Eq.(\ref{res1}), 
the finite coupling $E_1^{(r)}(\lambda)$ can be utilized to remove {\it quark-mass independent} short distance physics contributions $\sim \Delta_0$ from loop 
diagrams contributing to this form factor. Implementing this constraint one finds
\begin{eqnarray}
E_1^{(r)}(\lambda) & = &
	-\frac{c_AM_N^2}{324\pi^2F_{\pi}^2}\left[36g_A-139g_1+3(35g_1-9g_A)
	\ln{\left(\frac{2\Delta_0}{\lambda}\right)}\right]. \label{Echoice}
\end{eqnarray}
We emphasize again that this special choice for $E_1^{(r)}(\lambda)$ does not lead to observable consequences\footnote{This construction 
ensures that contributions from loops involving $\Delta(1232)$ as an intermediate state get suppressed once the mass of $\Delta(1232)$ gets larger. For
a fixed value of the mass of $\Delta$(1232) this decoupling-construction is not necessary.} in the final result, except for
changing the numerical values of the couplings $b_1$ and $D_1^{(r)}(\lambda)$ by taking away short distance physics $\sim \Delta_0$ 
arising in the loop 
integrals. Unfortunately, at ${\cal O}(\epsilon^3)$ in SSE we cannot\footnote{At ${\cal O}(\epsilon^4)$ in SSE we will be able to 
separate contributions from $b_1$ and $D_1^{(r)}(\lambda)$ via differences in the quark-mass dependence \cite{GH}.} 
separate the three independent couplings $b_1,\,b_6$ and $D_1^{(r)}(\lambda)$, 
as we only encounter the two linearly independent combinations $A(\lambda),\, B(\lambda)$ which we fit to experimental input in 
section \ref{sec:res}. We are therefore postponing a discussion of the 
size of the individual couplings to a future communication \cite{GH}.

Finally we would like to comment on the constant $C_s$ in Eq.(\ref{res2}). To ${\cal O}(\epsilon^3)$ in non-relativistic SSE 
all counter terms---i.e. all short distance physics contributions---only appear at $q^2=0$, cf. Eqs.(\ref{res1}-\ref{res3}). All radii of the $G_i(q^2)$ 
form factors therefore arise as {\em pure loop effects from the chiral pion-dynamics} at this order. While it is somewhat expected that the 
dominant parts
of these isovector $N\Delta$-transition radii arise from the pion-cloud, it is also known---for example from calculations of 
the isovector nucleon form factors (see ref.\cite{BFHM})---that short distance contributions in such radii cannot be completely 
neglected. A short distance contribution to the $r_2$-radius of the
$N\Delta$-transition form factor $G_2(q^2)$
like $C_s$ could, for example, arise from the non-relativistic reduction of the ${\cal O}(\epsilon^5)$ SSE Lagrangean
\begin{eqnarray}
\mathcal{L}_{N\Delta}^{(5)}&\sim&C_s\bar{\psi}_\mu^{i}\gamma_5\frac{1}{2}\left(\tau^{i}\left[D^2,f^{\mu\nu}_+\right]\right)D_\nu\psi_N + h.c. \label{defC} .
\end{eqnarray}
While this contribution formally is suppressed by two orders in the SSE expansion, we will argue in section \ref{fit2} that the inclusion of
such a short distance coupling is crucial for a comparison with phenomenology. We note that the contribution of Eq.(\ref{defC}) to the form factor
$G_2(q^2)$ was not considered in ref.\cite{GHKP} and constitutes our main change in terms of formalism compared to those previous results.
We note that such radius terms do exist for all
form factors, also for the radii of $G_1$ or $G_3$. We have checked that an
inclusion of such terms in $G_1$ or $G_3$ does not lead to significant changes
of the best fit curves, indicating that such terms in $G_1$ and $G_3$ do behave as
small higher order corrections as suggested by the power counting.
At the end of section \ref{chiral} we will also discuss the chiral extrapolation of lattice results for the isovector $N\Delta$-transition form factors
in the multipole basis at $q^2=0$. As the quark-mass dependence of 
$\mathcal{G}_M^*(0)$, $\mathcal{G}_E^*(0)$, $\mathcal{G}_C^*(0)$ is rather
involved (see e.g. the definition equations (\ref{con1}-\ref{con3})), the specific form of their chiral extrapolation can only be given
numerically. However, the leading quark-mass dependence of the $G_i(q^2),\,i=1\ldots 3$ form factors at $q^2=0$ can be given in a closed 
form:
\begin{eqnarray}
G_1(0) & = &
-\frac{1}{2}b_1-D_1(\lambda)\frac{\Delta_0}{4M_N}+\frac{ic_Ag_A\Delta_0M_N}{24\pi
	F_{\pi}^2}\nonumber\\&&
	+\frac{c_AM_Nm_{\pi}^2}{576\Delta_0\pi^2F_{\pi}^2}\left[9g_A\left(6-3\pi^2+4i\pi-
	4\ln{\left(\frac{2\Delta_0}{m_{\pi}}\right)}\left(1+2i\pi-\ln{\left(\frac{2\Delta_0}{m_{\pi}}\right)}\right)\right)
	\right.\nonumber\\&&\left.
	-5g_1\left(10+3\pi^2+4
	\ln{\left(\frac{2\Delta_0}{m_{\pi}}\right)}+12\ln^2{\left(\frac{2\Delta_0}{m_{\pi}}\right)}\right)\right]
	+\frac{c_A(5g_1+9g_A)M_N
	m_{\pi}^3}{216\Delta_0^2\pi F_{\pi}^2}+... , \label{G1m}\\
G_2(0) &  = &2b_6-D_1(\lambda)-
	\frac{c_AM_N^2}{162\pi^2F_{\pi}^2}\left[5g_1\left(1+3
	\ln{\left(\frac{2\Delta_0}{\lambda}\right)}\right)+9g_A\left(1+3
	\left(i\pi-\ln{\left(\frac{2\Delta_0}{\lambda}\right)}\right)\right)\right]\nonumber\\&&
	+\frac{c_AM_N^2m_{\pi}^2}{144\Delta_0^2\pi^2F_{\pi}^2}\left[9g_A\left(10-3\pi(\pi-4i)-
	4\ln{\left(\frac{2\Delta_0}{m_{\pi}}\right)}\left(3+2i\pi-\ln{\left(\frac{2\Delta_0}{m_{\pi}}\right)}\right)
	\right)\right.\nonumber\\&&
	\left.-5g_1\left(10+\pi^2-12
	\ln{\left(\frac{2\Delta_0}{m_{\pi}^2}\right)}
	+4\ln^2{\left(\frac{2\Delta_0}{m_{\pi}}\right)}
	\right)\right]
	+\frac{M_N^2c_A(5g_1+9g_A)m_{\pi}^3}{27\Delta_0^3\pi F_{\pi}^2}+... ,\label{G2m} \\
G_3(0) & =
	&\frac{c_AM_N^2}{648\pi^2F_{\pi}^2}\left[85g_1-9g_A(17+6i\pi)+6(9g_A-5g_1)
	\ln{\left(\frac{2\Delta_0}{m_{\pi}}\right)}\right]\nonumber\\&&
	-\frac{c_A(5g_1+9g_A)M_N^2m_{\pi}}{36\Delta_0\pi F_{\pi}^2}+\frac{c_AM_N^2m_{\pi}^2}{288\Delta_0^2\pi^2F_{\pi}^2}
	\Bigg[
	9g_A\Bigg(-22+9\pi^2-20i\pi+\nonumber\\&&
	4\ln{\left(\frac{2\Delta_0}{m_{\pi}}\right)}\left(
	5+6i\pi-3\ln{\left(\frac{2\Delta_0}{m_{\pi}}\right)}\right)\Bigg)+5g_1\left(
	22+3\pi^2+4\ln{\left(\frac{2\Delta_0}{m_{\pi}}\right)}\left(-5+3\ln{\left(\frac{2\Delta_0}{m_{\pi}}\right)
	}\right)\right)\Bigg]\nonumber\\&&
	-\frac{c_A(5g_1+9g_A)M_N^2m_{\pi}^3}{27\Delta_0^3\pi
F_{\pi}^2}+...\, \label{chex}.
\end{eqnarray} 
We observe that the leading quark-mass $m_q$ dependence\footnote{Here we assume the validity of the Gell-Mann, Oakes, Renner 
relation \cite{GOR} given in Eq.(\ref{GOR}) in order to convert the $m_\pi$ dependence into the $m_q$ dependence.} both for $G_1(0)$ and for $G_2(0)$ is 
linear in $m_q$, whereas the leading non-analytic quark-mass behaviour is proportional to $m_q\log m_q$, both for the real and for the
imaginary parts. On the other hand, $G_3(0)$ displays a chiral singularity $\sim\log m_q$ near the chiral limit, which will also appear in
the Coulomb quadrupole form factor $\mathcal{G}_C^*(0)$ in section \ref{chiral}. Finally, we observe that to ${\cal O}(\epsilon^3)$ short distance
effects arising from the loop integrals could be removed in
Eq.(\ref{G1m}) from the {\em real} part of $G_1(0)$ 
via the  choice for $E_1^{(r)}(\lambda)$ given in Eq.(\ref{Echoice}),
whereas the real part of $G_2(0)$
in Eq.(\ref{G2m}) and all imaginary parts are still affected by
quark-mass independent short distance physics generated by the loop diagrams of Fig.\ref{fig:feyn}. This 
nuisance can only be remedied at the next order ${\cal O}(\epsilon^4)$ \cite{GH}. In the present ${\cal O}(\epsilon^3)$ SSE calculation this
situation is partly responsible for the rather large fit-values we will obtain for $A(\lambda),\,B(\lambda)$ in the next section.
 
\section{Discussion of the results}
\label{sec:res} 
\subsection{Fit 1: Comparison to previous ${\cal O}(\epsilon^3)$ SSE results} \label{fit1}

\begin{figure}
\begin{center}
\includegraphics[width=12cm,clip]{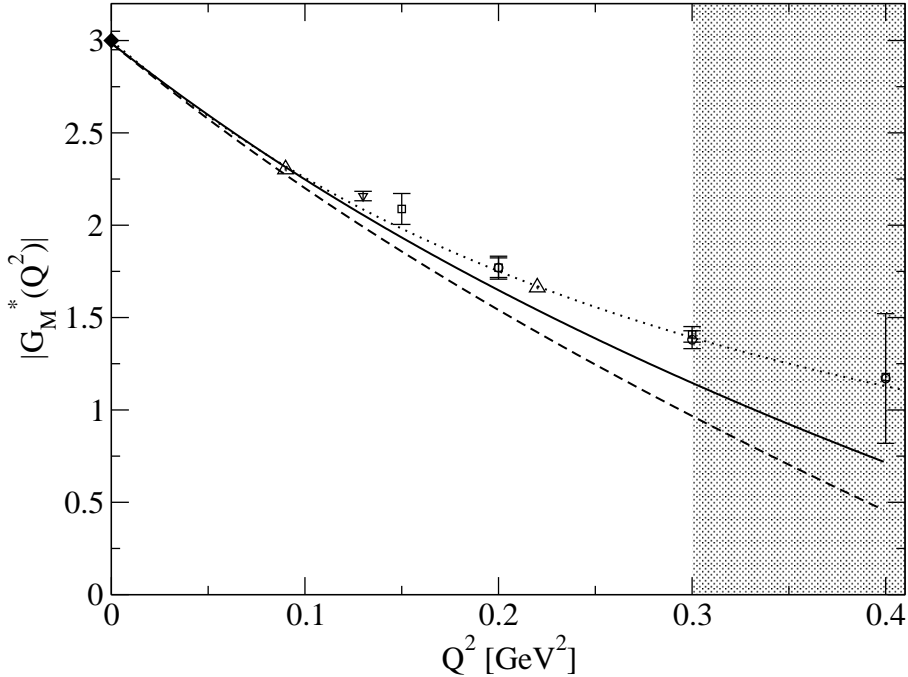}
\caption{The absolute value of the magnetic $N\Delta$-transition form factor $\mathcal{G}_M^{*Ash}(Q^2)$ in the convention of 
Eq.(\ref{eq:ash}) to ${\cal O}(\epsilon^3)$ in non-relativistic SSE. Solid line:
Fit 2. Dashed line: Fit 1. Dotted line: MAID parametrization \cite{param}. Experimental data from \cite{Beck} (diamond),
 \cite{Stein} (triangle up), \cite{rean} (triangle down), \cite{Baetzner} (square) and \cite{Bartel} (circle).}
\label{GM}
\end{center}
\end{figure}

The strict $\mathcal{O}(\epsilon^3)$ results of ref.\cite{GHKP} can be obtained from Eqs.(\ref{res1}-\ref{res3}) by setting $C_s\equiv 0$. For the
numerical values of the input parameters we do not follow that reference but instead we utilize the updated values for the couplings given in 
table \ref{tab:param}.
In Fit 1 we determine the two unknown parameters $A(\lambda)$ and $B(\lambda)$ 
by inserting Eqs.(\ref{res1})-(\ref{res3}) into
Eqs.(\ref{con1})-(\ref{con3}) and fit \\
a) to the experimental data for $\left|\mathcal{G}_M^{*Ash}(Q^2)\right|$  shown in figure \ref{GM}
for momentum transfers of
$Q^2<0.2$ GeV$^2$ and simultaneously \\
b) to the experimental value for EMR$(0)$ of ref.\cite{Beck} utilizing
Eq.(\ref{CMR}).\\ 
The resulting values\footnote{We discuss the numerical size of the two parameters in section \ref{fit2}.} of this Fit 1 
are given in table \ref{fit} for a regularization scale of $\lambda=1$ GeV. 
\begin{table}
\begin{center}
\begin{tabular}{|c|c|c|c|c|c|c|c|}
\hline
Parameter & $g_A$ & $c_A$ & $g_1$ & $M_N$ [GeV] & $M_{\Delta}$ [GeV] & $
m_{\pi}$ [GeV] & $F_{\pi}$ [GeV] \\
Value & 1.26 & 1.5 & 2.8 & 0.939 & 1.210 & 0.14 & 0.0924\\
\hline
\end{tabular}
\caption{The input parameters for our calculation. The nucleon properties $g_A$ and $M_N$ are taken from \cite{PDG}. For the definition of
the mass of $\Delta$(1232) we are utilizing the T-matrix definition of \cite{PDG}, leading to a (real part) mass of $1210$ MeV. The coupling 
$g_1$ has recently been determined in \cite{QCDSF}, whereas the value for $c_A$ is derived in Appendix \ref{app:ca}.}
\label{tab:param}
\end{center}
\end{table}

\begin{figure}
\begin{center}
\includegraphics[width=12cm,clip]{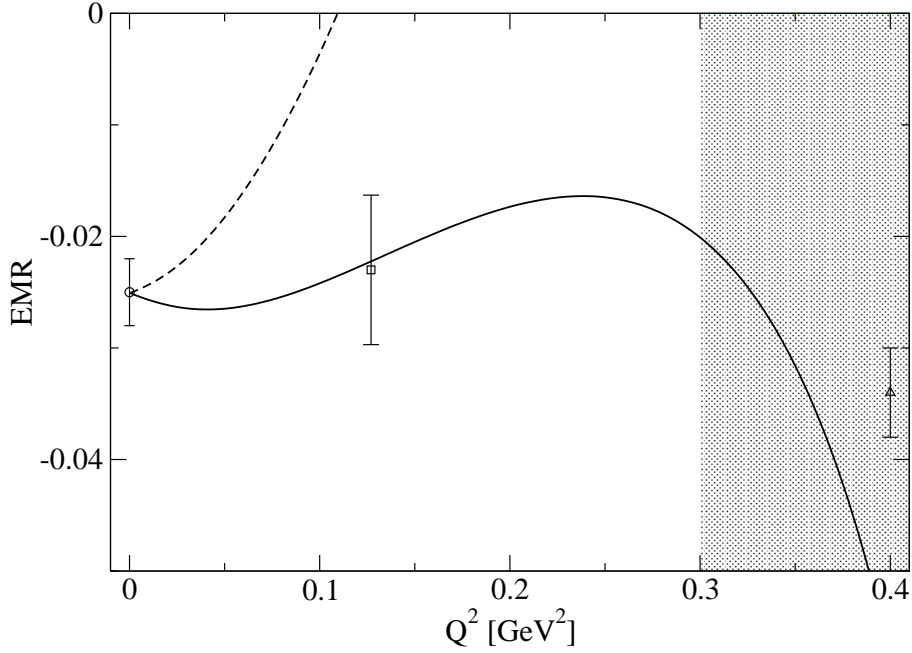}
\caption{EMR$(Q^2)$ to ${\cal O}(\epsilon^3)$ in non-relativistic SSE. Dashed line: Fit 1.
Solid line: Fit 2. Experimental data at 
the real photon point from MAMI \cite{Beck}, at $Q^2=0.127$ GeV$^2$ from OOPS \cite{OOPS} and 
at $Q^2=0.4$ GeV$^2$ from CLAS \cite{CLAS}.}
\label{fig:EMR}
\end{center}
\end{figure}

\begin{figure}
\begin{center}
\includegraphics[width=12cm,clip]{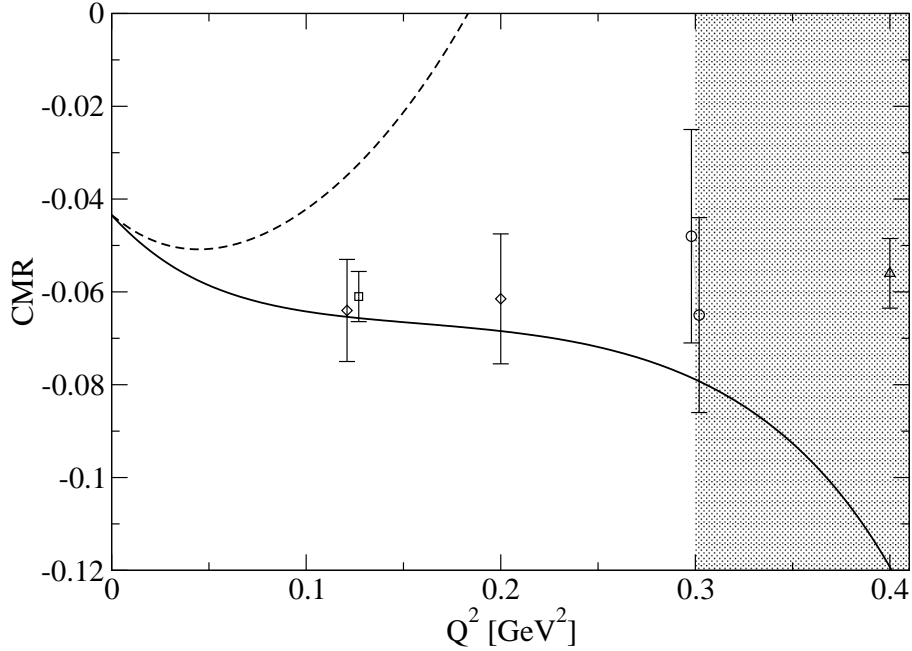}
\caption{CMR$(Q^2)$ to ${\cal O}(\epsilon^3)$ in non-relativistic SSE. Dashed line: Fit 1.
Solid line: Fit 2. The data-points shown are from refs. \cite{MAMI} (diamonds), \cite{Bonn} (circles) and \cite{CLAS} (triangle).}
\label{fig:CMR}
\end{center}
\end{figure}

\begin{figure}
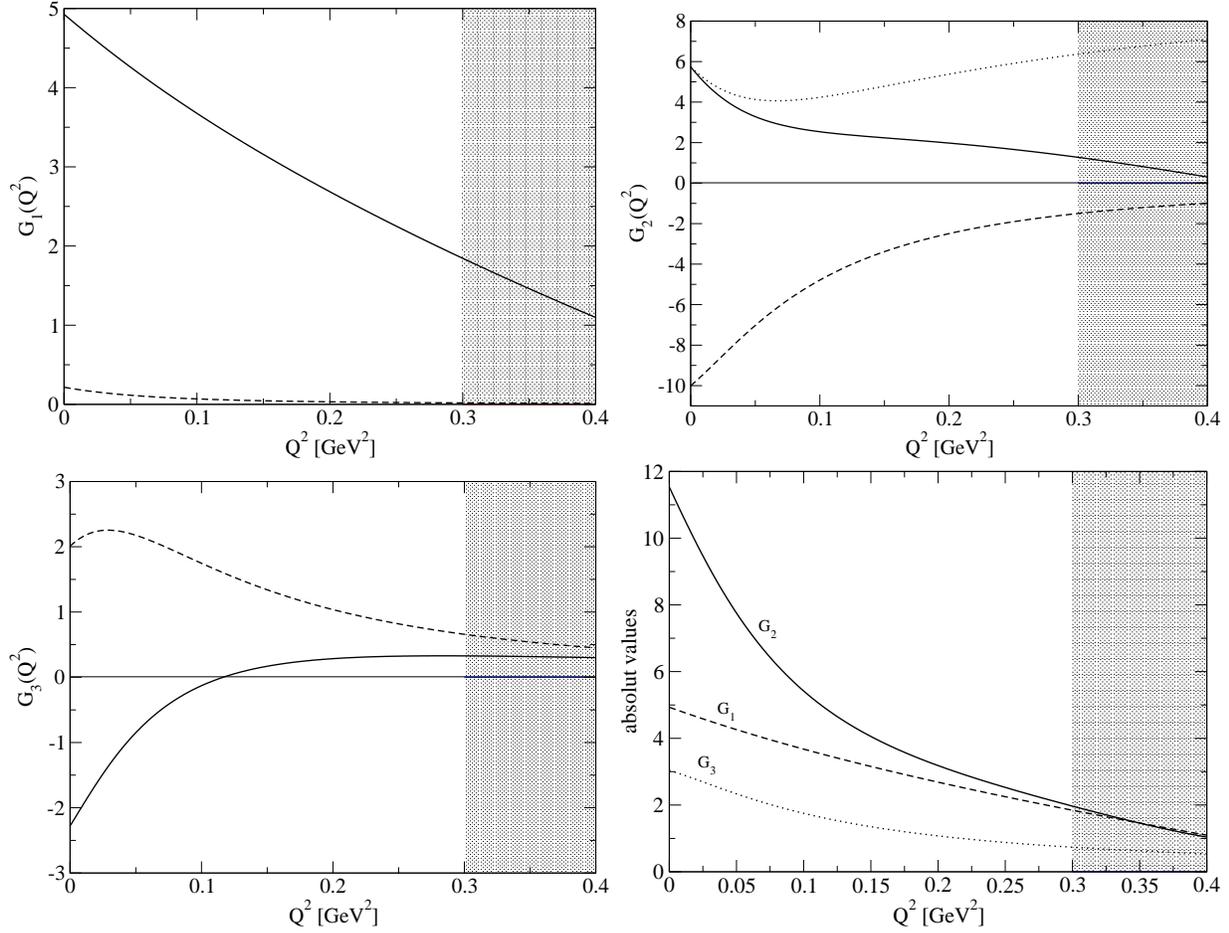

\begin{center}
\includegraphics[width=8cm,clip]{ReImG1.eps}
\includegraphics[width=8cm,clip]{ReImG2.eps}
\includegraphics[width=8cm,clip]{ReImG3.eps}
\includegraphics[width=8cm,clip]{Abs.eps}
\caption{The form factors $G_1$, $G_2$ and $G_3$ defined in
Eq.(\ref{defff}). Solid lines show the real parts, dashed lines
imaginary parts. The dotted line in the plot for $G_2(Q^2)$ arises in Fit 1, whereas the solid line corresponds to Fit 2. The absolute value of 
$G_2(Q^2)$ shown in the panel on the lower right corresponds to Fit 2.}
\label{fig:Gi}
\end{center}
\end{figure}

\begin{table}
\begin{center}
\begin{tabular}{|c|c|c|c|}
\hline
Parameter & $A$(1GeV) & $B$(1GeV) & $C_s$ [GeV$^{-2}$] \\
Fit I &10.5 & 15.4& 0 (fixed) \\
Fit II &10.5 & 15.4 &-17.0\\ 
\hline
\end{tabular}
\end{center}
\caption{The values for the unknown parameters obtained by fitting our results to the
experimental data for
$\left|\mathcal{G}_M^{*Ash}(Q^2<0.2GeV^2)\right|$ and
EMR$(0)$ at a regularization scale of $\lambda=1$ GeV. In Fit I we set
$C_s=0$. We note that the values for A and B do not change significantly
between Fit 1 and Fit 2.} 
\label{fit}
\end{table}
The dashed curve in Figure \ref{GM} shows, that this procedure leads to a satisfying description
of
$\left|\mathcal{G}_M^{*Ash}(Q^2)\right|$ up to $Q^2\approx0.2$ GeV$^2$. We note that the dotted
curve also shown in that figure is the parametrization of the MAID result
\cite{param}, which takes into account the fact that this form factor
is falling faster than the dipole by inserting an extra exponential function:
\begin{eqnarray}
\left|\mathcal{G}_M^{*Ash}(Q^2)\right|& = &
\frac{3}{\left(1+\frac{Q^2}{0.71\textnormal{GeV}^2}\right)^2}\exp{\left(-0.21\frac{Q^2}{\textnormal{GeV}^2}\right)}.
\label{dip}
\end{eqnarray}
However, while we obtain a reasonable $Q^2$-dependence for the magnetic $N\Delta$-transition form factor up to 
$Q^2\approx 0.2$ GeV$^2$, we only get the right value of EMR($Q^2$) at
the photon point $Q^2=0$, while the $Q^2$-dependence of this ratio 
is far off the experimental data. This can be seen from the dashed curve in Fig.\ref{fig:EMR}. A similarly non-satisfying picture results
for the $Q^2$-dependence of the CMR-ratio, see Fig.\ref{fig:CMR}. We have analysed the reason for these
early breakdowns in $Q^2$ of the ${\cal O}(\epsilon^3)$ SSE
calculation of ref.\cite{GHKP}. According to our new analysis
presented here these small ratios are
very sensitive to the exact form of the $G_2(Q^2)$ form factor.  As it
can be seen from the dotted curve of $G_2(Q^2)$ in 
Fig.\ref{fig:Gi}, in Fit 1 the momentum transfer dependence of the 
real part of $G_2$ has an (unphysical) turning point already at rather low $Q^2$. It begins to rise again above
$Q^2\approx0.05$ GeV$^2$. This
``unnatural''\footnote{We consider this behaviour
to be unphysical, because we expect the momentum dependence of a baryon form factor in an
effective theory to decrease in magnitude in the momentum range
$0.1$GeV$^2 < Q^2 << \Lambda_{\chi}^2 \approx 1$GeV$^2$ when the
resolution is increased. For further examples of this observed
behaviour in the case of nucleon form factors we point to refs.\cite{BFHM,QCDSFff}
} behaviour is an indication that important physics is not
included in the SSE-calculation at the order we are working. Our
analysis shows that it is this ``unphysical'' behaviour of the $G_2(q^2)$ form
factor which is responsible for the poor results of the $Q^2$-dependence in EMR$(Q^2)$ and CMR$(Q^2)$ in Fit 1. 
In the next section we will present a remedy for this breakdown. In conclusion
we must say that for state-of-the art coupling constants\footnote{The curves given in ref.\cite{GHKP} look slightly different from the ones given
here as Fit 1 due to updated numerical values for the input parameters used here.} as given in table \ref{tab:param} the non-relativistic
${\cal O}(\epsilon^3)$ SSE calculation for the (small)
electric and coulomb $N\Delta$-quadrupole transition form factors of ref.\cite{GHKP} is only valid up for $Q^2<0.05$ GeV$^2$, while 
the (much larger) magnetic $N\Delta$-transition form factor is described well with the results of ref.\cite{GHKP} over a larger range in $Q^2$.

\subsection{Fit 2: Revised ${\cal O}(\epsilon^3)$ SSE analysis} \label{fit2}
  
The early breakdown of the ${\cal O}(\epsilon^3)$ SSE calculation of ref.\cite{GHKP} discussed in the previous section can be
overcome by introducing the parameter $C_s$ in Eq.(\ref{res2}). As discussed in section \ref{sec:calc} such a term formally arises from
higher order couplings in the ${\cal O}(\epsilon^5)$ Lagrangean like the one displayed in Eq.(\ref{defC}). Physically, this term amounts
to a (small) short distance correction in the radius of the form
factor $G_2(Q^2)$, which in Fit 1 is given solely by pion-loop
contributions\footnote{An important check for the validity of our
interpretation of $C_S$ can be provided by an
$\mathcal{O}(\epsilon^4)$ analysis \cite{GH} which should lead to the same
conclusion regarding the size of short distance effects in $r_2$.}.
One may wonder, why such a contribution, which formally is of much
higher order in the perturbative  chiral calculation, suddenly should play such a
prominent role. However, we have to point out that
the (rather small) electric $\mathcal{G}_E^*(Q^2)$ form factor is very sensitive to the
$N\Delta$-transition form factor $G_2(Q^2)$. The size of this quadrupole form factor is at the percent level of the dominant  
magnetic $\mathcal{G}_M^*(Q^2)$ transition form factor---small changes in $G_2(Q^2)$ are therefore disproportionally magnified 
when looking at EMR$(Q^2)$.
In the following we will explicitly include the
radius term in Fit 2, which now has three unknown parameters 
$A(\lambda), B(\lambda)$ and
$C_s$. Utilizing the same input parameters as in Fit 1 (see table \ref{tab:param}) we insert Eqs.(\ref{res1})-(\ref{res3}) into
Eqs.(\ref{con1})-(\ref{con3}) and fit again \\
a) to the same experimental data for $\left|\mathcal{G}_M^{*Ash}(Q^2)\right|$  shown in Fig.\ref{GM}
at momentum transfers of
$Q^2<0.2$ GeV$^2$ and simultaneously \\
b) to the experimental value for EMR$(0)$ reported in ref.\cite{Beck}
utilizing Eq.(\ref{CMR}). \\
We note that neither in Fit 1 of the previous section nor in
the new Fit 2 we have fit to the CMR$(q^2)$ data of Fig.\ref{fig:CMR}. In both fits the resulting curves are a prediction. 
The results for the three parameters of this new Fit 2 are given in table \ref{fit}. First we notice that the central values for the parameters 
$A(1$ GeV$), B(1$ GeV$)$ have not changed significantly. While the numerical value for the new parameter $C_s$ is quite large, it 
actually only amounts to a small correction of $0.21$ fm in the $r_2$ radius, in agreement with the expectation from the chiral counting:
\begin{eqnarray}
r_{2,\textnormal{Re}}:\,1.57 \,\textnormal{fm (Fit
1)}&\rightarrow&1.78 \,\textnormal{fm (Fit 2)}.
\label{r2shift}
\end{eqnarray}
This small correction in the radius of the $G_2(Q^2)$ form factor leads to a much more physical behaviour\footnote{We note that we do not 
expect that the zero-crossing in the real part of the form factor $G_3(Q^2)$ near $Q^2=0.1$ GeV$^2$ in Fig.\ref{fig:Gi} corresponds to a 
physical behaviour. However, the size of $Re\left[G_3(Q^2)\right]$ for $Q^2>0.1$ GeV$^2$ is so small that this effect does not affect our
results in any significant way. For completeness we note that at
${\cal O}(\epsilon^4)$ there is a counter term in the SSE Lagrangean which
will lead to a momentum-independent overall shift in $Re\left[G_3(Q^2)\right]$ which should correct this presumed artifact \cite{GH}.} 
in the real part of $G_2$ for $Q^2<0.4$ GeV$^2$, as can clearly be seen from the solid curve in Fig.\ref{fig:Gi}. The resulting changes in the
momentum-dependence of EMR and CMR as shown in the solid curves of
Figs.\ref{fig:EMR}, \ref{fig:CMR} are quite astonishing. The small 
change in the radius of $G_2$ has lead to agreement with experimental data both for EMR and CMR up to a four-momentum transfer 
squared of $Q^2\approx 0.3$ GeV$^2$. We note again that none of the experimental data points at {\em finite $Q^2$} in EMR or CMR have been used 
as input for the determination of the
fit-parameters. The significant change of
the Quadrupole form factors caused by the inclusion of the radius term
which is formally of higher order proves that it underestimated by naive
power counting. At the same time the resulting good accordance with phenomenology shows, that this
term includes relevant physics into our calculation.
These two observations\footnote{Experience in
ChEFT teaches that one has to be particularly careful in those
one-loop calculations which are finite without any short distance
couplings required to absorb possible divergences. Examples include
the electric polarizability $\alpha_E$ to $
\mathcal{O}(\epsilon^3)$ in SSE (see ref.\cite{Robert}) or the
contributions beyond the finite one-loop result of the process $\gamma\gamma\rightarrow \pi^0 \pi^0$ in refs.\cite{gg1,gg2}.} constitute our justification for the inclusion
of the coupling $C_S$.
For completeness we also note that the resulting absolute value of 
$\mathcal{G}_M^*(Q^2)$ of Fit 2 is
now also in decent agreement with the experimental data up to
$Q^2\approx 0.2\ldots 0.3$ GeV$^2$. We therefore conclude that in Fit
2, after the radius correction in $G_2(Q^2)$ has been inserted, it is now the 
insufficient momentum-dependence of $\mathcal{G}_M^*(Q^2)$ {\em above} $Q^2\approx0.25$ GeV$^2$ that sets the limit in $Q^2$ for the new 
non-relativistic 
${\cal O}(\epsilon^3)$ SSE result of Fit 2 presented here. We
indicate this limitation by the gray-shaded bands in
Figs.\ref{GM}-\ref{fig:CMR} and 
will explore this point further in section \ref{breakdown}

\begin{figure}
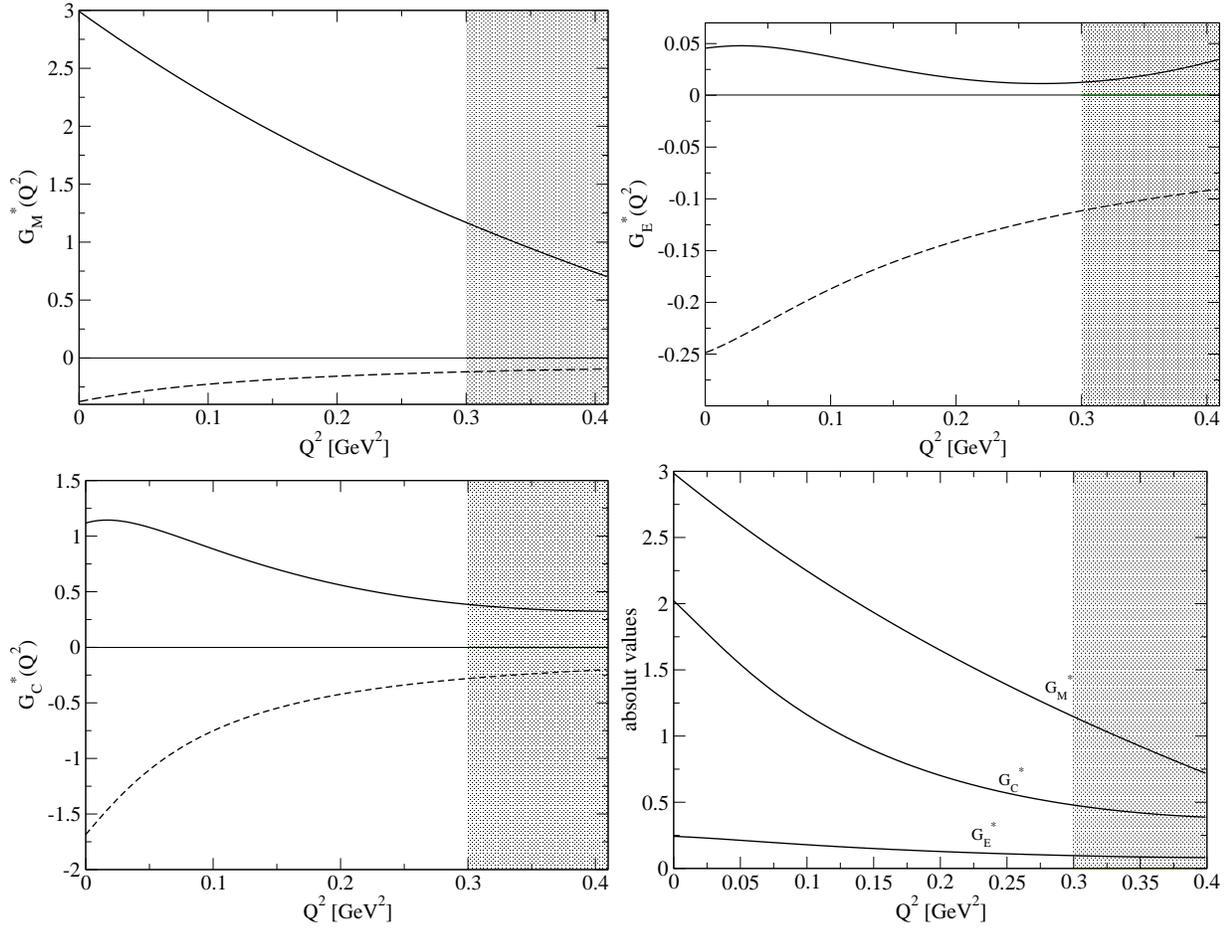

\begin{center}
\includegraphics[width=8cm,clip]{ReImGM.eps}
\includegraphics[width=8cm,clip]{ReImGE.eps}
\includegraphics[width=8cm,clip]{ReImGC.eps}
\includegraphics[width=8cm,clip]{Abs3.eps}
\caption{
$\mathcal{O}(\epsilon^3)$ SSE results of Fit 2 for the $N\Delta$-transition form factors in the multipole-basis. Solid
lines: Real parts. Dashed lines: Imaginary parts.} \label{fig:main}
\end{center}
\end{figure}

In table \ref{tab:radii} we present the results of Fit 2 for all six $N\Delta$-transition form factors discussed in this work, both at $Q^2=0$ and for 
their (complex) radii, defined as:
\begin{eqnarray}
G_i(Q^2) & = &
\textnormal{Re}[G_i(0)]\left[1-\frac{1}{6}r_{i,\textnormal{Re}}^2Q^2+...\right]+
i\,\textnormal{Im}[G_i(0)]\left[1-\frac{1}{6}r_{i,\textnormal{Im}}^2Q^2+...\right].
\end{eqnarray}
\begin{table}
\begin{center}
\begin{tabular}{|c|cc|cc|cc|}
\hline
  & $\textnormal{Re}[G_i(0)]$ & $r_{i,\textnormal{Re}}^2$
[fm$^{2}$] & $\textnormal{Im}[G_i(0)]$
 & $r_{i,\textnormal{Im}}^2$ [fm$^{2}$] &
 $\left|G_i(0)\right|$ & $r_{i,\textnormal{Abs}}^2$ [fm$^{2}$] \\
\hline
$G_1$ & 4.95& 0.679 &0.216 & 3.20 & 4.96 & 0.678 \\
$G_2$ & 5.85& 3.15 & -10.0 & 1.28 & 11.6 & 1.73 \\
$G_3$ & -2.28& 3.39 & 2.01 & -2.26& 3.04 & 0.907 \\
\hline
$\mathcal{G}_M^*$ & 2.98 & 0.627 &-0.377 & 1.36 & 3.00 & 0.630 \\
$\mathcal{G}_E^*$ & 0.0441&-0.836 &-0.249 & 0.422& 0.253 & 0.388\\
$\mathcal{G}_C^*$ & 1.10& -0.729 &-1.68 & 1.90 & 2.01 & 1.10 \\
\hline
\end{tabular}
\caption{The values at $Q^2=0$ and the radii of the two sets of form
factors used in this work obtained in Fit 2. }
\label{tab:radii}
\end{center}
\end{table}
It is interesting to note that the real parts of the radii of both the electric and the coulomb $N\Delta$-transition form factors are negative! This can
also be observed in Fig.\ref{fig:main}: The quadrupole $N\Delta$-transition form factors definitely do not behave like dipoles in the low 
$Q^2$-region, they look different both from the Sachs form factors of the nucleon and from the common parametrization for 
$\mathcal{G}_M^*(Q^2)$ of Eq.(\ref{dip}).

Table \ref{tab:radii} and  Fig.\ref{fig:main} constitute the central results of our analysis. They make clear, that the non-trivial
$Q^2$-dependence of EMR$(Q^2)$ and CMR$(Q^2)$ observed in Figs. \ref{fig:EMR}, \ref{fig:CMR} arises from the quadrupole transition 
form factors, which should therefore be studied independently of $\mathcal{G}_M^*(Q^2)$. Finally we want to comment on the size of the 
short distance contributions ``$sd$'' parametrized via $A(1$ GeV$),\,
B(1$ GeV$)$ versus the long distance contributions from the pion-cloud ``$pc$''. Despite the
large values for these combinations of LECs (see
Eqs.(\ref{Al},\ref{Bl})), the $N\Delta$-transition form factors are not
completely dominated by short distance 
physics \footnote{Despite the seemingly large values for A and B as
given in
table \ref{fit}, the size of the short distance contributions in
$\mathcal{G}^*_M(0)$, $\mathcal{G}^*_E(0)$ and $\mathcal{G}_C^*(0)$
is natural as expected (see Eqs.(\ref{ps1}-\ref{ps2})).}---clear signatures of chiral dynamics are visible.
At a scale of $\lambda=1$ GeV one obtains (at the photon point):
\begin{eqnarray}
\textnormal{Re}[\mathcal{G}_M^*(0)] |_{\lambda=1\,\textnormal{GeV}}& = &
\left.-1.06\right|_{pc}+\left.4.04\right|_{sd},
\label{ps1}\\
\textnormal{Re}[\mathcal{G}_E^*(0)] |_{\lambda=1\,\textnormal{GeV}} & = & \left.0.155\right|_{pc}-\left.0.110\right|_{sd},\\
\textnormal{Re}[\mathcal{G}_C^*(0)] |_{\lambda=1\,\textnormal{GeV}} & = & \left.1.47\right|_{pc}-\left.0.365\right|_{sd}.
\label{ps2}
\end{eqnarray}
We note that such a separation into short and long range physics is obviously scale-dependent. However, at a much lower regularization scale of $\lambda=600$ MeV we have 
checked that one arrives at the same pattern. Analyzing Eqs.(\ref{ps1}-\ref{ps2}) we conclude that the magnetic $N\Delta$-transition is dominated by short distance physics. Its strength
is ${\it reduced}$ by $\approx 40\%$ due to pion-cloud effects in the
magnetic $N\Delta$-transition. This result is very similar to the situation in the isovector magnetic moment of the 
nucleon (see the discussion in ref.\cite{HW}), both in sign and in
magnitude!\\
 While the pion-cloud and the short distance physics are of the same
magnitude but of opposite sign for the very small electric quadrupole transition moment, the coulomb quadrupole moment in our non-relativistic
${\cal O}(\epsilon^3)$ SSE analysis is dominated by the chiral dynamics of the pion-cloud. We note that the small resulting
value for CMR$(Q^2)$ is of purely kinematical origin (see Eq.(\ref{CMR})), whereas the electric quadrupole form factor in the Jones-Scadron 
conventions used in this work is intrinsically small relative to the magnetic M1 transition form factor. We will report in a future communication
\cite{GH} whether this pattern Eqs.(\ref{ps1}-\ref{ps2}) will hold also at next-to-leading one-loop order (i.e. ${\cal O}(\epsilon^4)$) in SSE, where we also hope to project
out individual values for three leading $\gamma N\Delta$-couplings $b_1,\,b_6$ and $D_1^{(r)}(\lambda)$ \cite{GH}. Before we discuss the 
quark-mass dependence of the form factors we first want to comment on the range of $Q^2$ in which the non-relativistic 
${\cal O}(\epsilon^3)$ SSE calculation seems applicable. 

\subsection{Range of applicability of the non-relativistic ${\cal O}(\epsilon^3)$ SSE calculation} \label{breakdown}

\begin{figure}
\begin{center}
\includegraphics[width=12cm,clip]{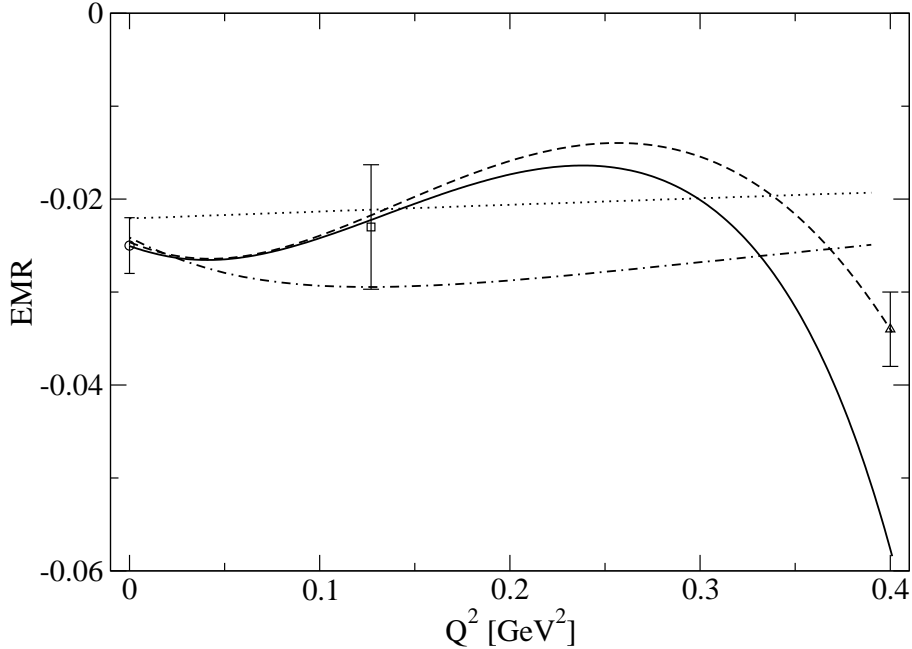}
\caption{EMR$(Q^2)$ at small momentum transfer. Solid line: ${\cal O}(\epsilon^3)$ SSE result of Fit 2. Dashed line: EMR$(Q^2)$ with 
parametrized $\mathcal{G}_M^*$ of Eq.(\ref{dip}) and the ${\cal O}(\epsilon^3)$ SSE result of Fit 2 for $\mathcal{G}_E^*$. The dotted
and dashed-dotted lines are results of the MAID2003 and DMT models \cite{DMTMAID}. For explanations on the data-points shown see 
Fig.\ref{fig:EMR}.}
\label{fig:EMRmod}
\end{center}
\end{figure}

\begin{figure}
\begin{center}
\includegraphics[width=12cm,clip]{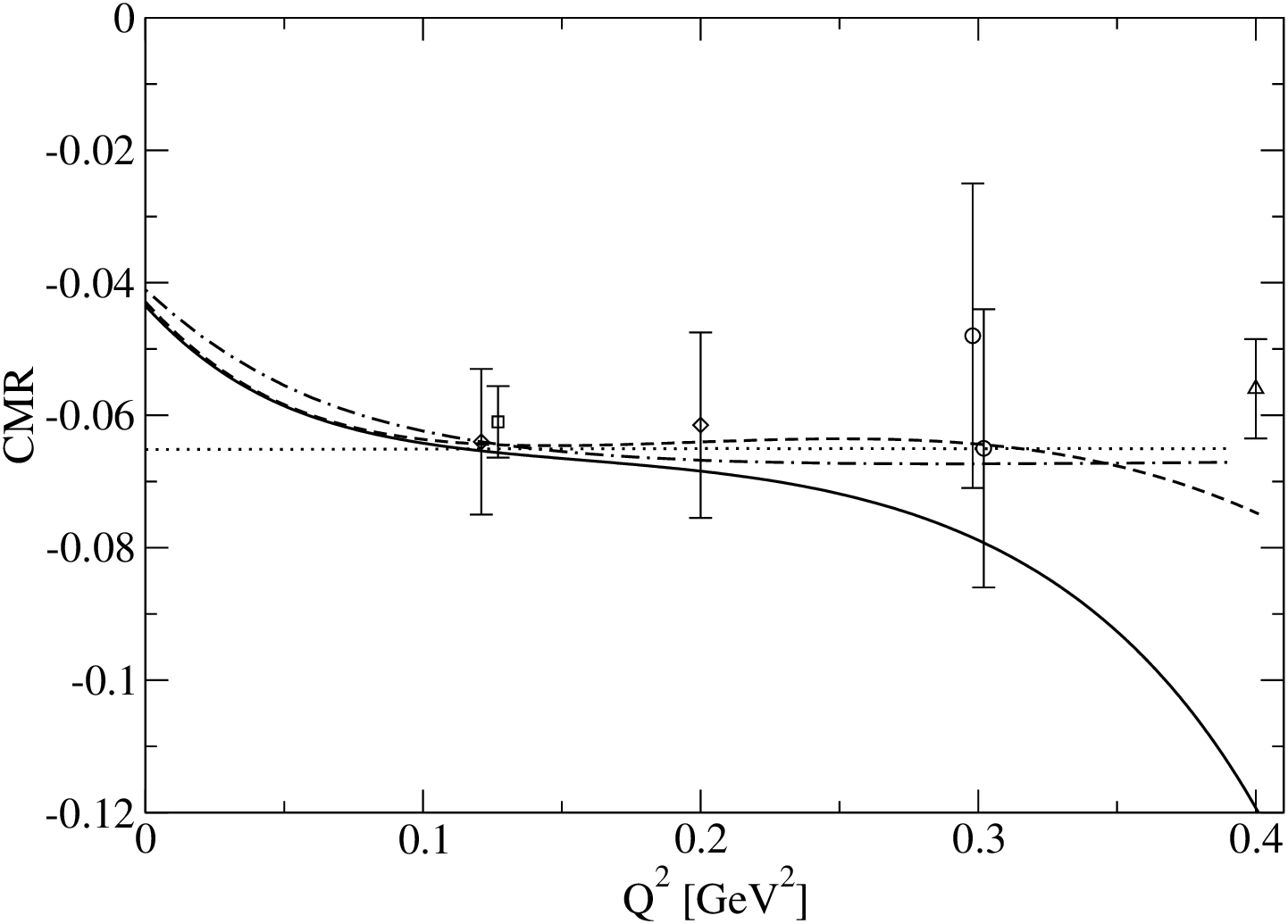}
\caption{CMR$(Q^2)$ at small momentum transfer. Solid line: ${\cal O}(\epsilon^3)$ SSE result of Fit 2. Dashed line: CMR$(Q^2)$ with 
parametrized $\mathcal{G}_M^*$ of Eq.(\ref{dip}) and the ${\cal O}(\epsilon^3)$ SSE result of Fit 2 for $\mathcal{G}_C^*$. The dotted
and dashed-dotted lines are results of the MAID2003 and DMT models \cite{DMTMAID}. For explanations on the data-points shown see 
Fig.\ref{fig:CMR}.} 
\label{fig:CMRmod}
\end{center}
\end{figure}

In non-relativistic ${\cal O}(\epsilon^3)$ SSE calculations the isovector Sachs form factors of the nucleon agree well with dispersion-theoretical 
results up to a four-momentum transfer of $Q^2\approx 0.3$ GeV$^2$
(see {\it e.g.} the discussion in ref.\cite{QCDSFff}). On the other hand, it is known that
covariant ChEFT calculations of baryon form factors usually do not
find enough curvature in the $Q^2$-dependence of such form factors beyond the radius term linear in 
$Q^2$ ({\it e.g.} see ref.\cite{KM}), due to a different organisation of the (perturbative) ChEFT series. We suspect that this is also the reason
why the $Q^2$-dependence of EMR and CMR reported recently in the covariant ChEFT calculation of ref.\cite{PV} is {\em markedly} different 
from the one 
presented here.\\
In our non-relativistic ${\cal O}(\epsilon^3)$ SSE analysis the limiting factor---as far as the $Q^2$-dependence is 
concerned---seems to be the deviation between the result of Fit 2 for $\mathcal{G}_M^*(Q^2)$ (solid curve in Fig.\ref{GM}) compared to the data (parametrized by the 
dotted curve). In order to demonstrate this we present EMR$(Q^2)$ and CMR$(Q^2)$ again in Figures \ref{fig:EMRmod}, 
\ref{fig:CMRmod}, now with $\mathcal{G}_M^*(Q^2)$ not given by our
result of Fit 2 but with the exponential parametrization of Eq.(\ref{dip}).
One can clearly observe that in both figures the agreement with the experimental results now extends to even larger values of $Q^2$, giving us confidence 
that the here calculated results for the electric and coulomb $N\Delta$ quadrupole-transition form factors---which are the quantities where 
the impact of chiral dynamics shows up most visibly (see the
discussion in section \ref{fit2})---have captured the relevant 
physics up to a momentum transfer $Q^2\approx 0.3$ GeV$^2$, similar to the situation for the isovector Sachs form factors of the nucleon
in non-relativistic ${\cal O}(\epsilon^3)$ SSE \cite{QCDSFff}. 
We also note that the slope of the SSE result for CMR at $Q^2<0.1$ GeV$^2$ is
highly dominated by $\pi N$ intermediate states (originating from
diagram (c) in Fig.\ref{fig:feyn}).
 The plateau in this ratio at higher momentum
transfer is due to a balance between this loop effect and short range
physics, with the larger $r_2$ radius of Fit 2 (see Eq.\ref{r2shift}) again being essential. We also observe in Fig.\ref{fig:CMRmod} that the DMT model of 
ref.\cite{DMTMAID} shows the same feature at low $Q^2$ as our SSE calculation. It is also interesting to note that the turnover in the 
$Q^2$-dependence of EMR near $Q^2\approx 0.25$ GeV$^2$ in Fig.\ref{fig:EMRmod} may not signal the breakdown of our approach
at this (already quite large) momentum transfer, but could indicate a
real structure effect connecting the OOPS and the CLAS results for EMR$(Q^2)$. In order
to decide this issue clearly the next-to-leading one-loop correction to our results has to be calculated \cite{GH}. 
Finally, we note again explicitly that the resulting dashed curves in Fig.\ref{fig:EMRmod}
and Fig.\ref{fig:CMRmod} have {\em not} been refit to the data points at finite
$Q^2$, despite their ``perfect`` agreement with the shown data.

\subsection{Chiral extrapolation of the $N\Delta$-transition form factors to ${\cal O}(\epsilon^3)$ in non-relativistic SSE} \label{chiral}

\begin{figure}
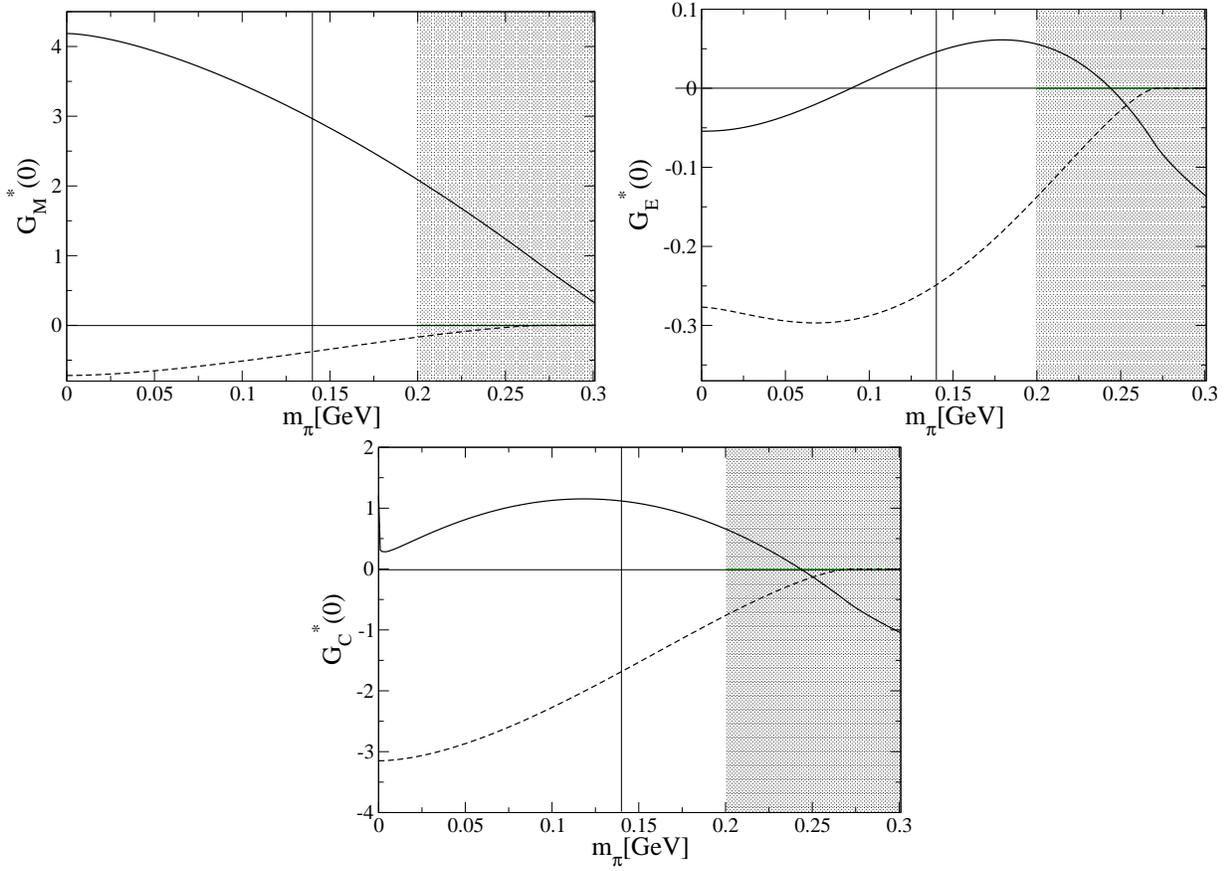

\begin{center}
\includegraphics[width=8cm,clip]{GMqm.eps}
\includegraphics[width=8cm,clip]{GEqm.eps}
\includegraphics[width=8cm,clip]{GCqm.eps}
\caption{The quark mass dependence of the $N\Delta$-transition form factors in the Jones-Scadron basis
according to non-relativistic $\mathcal{O}(\epsilon^3)$ SSE. Solid lines denote the real parts,
dashed lines the imaginary parts. We note that we have assumed here
that all masses appearing in the definitions of the Jones-Scadron
multipole form factors are taken at their values at the physical
point, in order to display {\em only} the intrinsic quark-mass
dependence of the transition form factors.}
\label{fig:qm}
\end{center}
\end{figure}

\begin{figure}
\begin{center}
\includegraphics[width=12cm,clip]{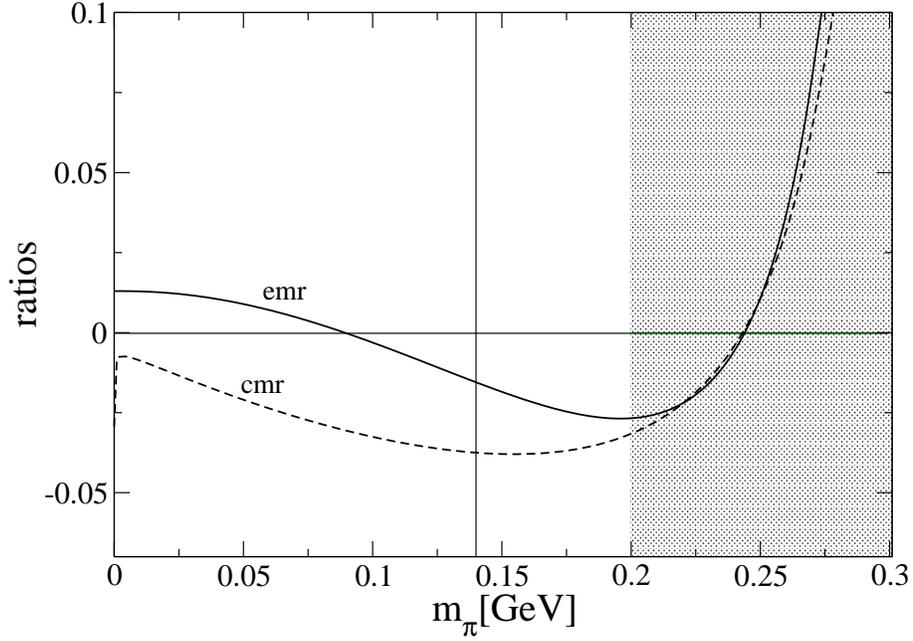}
\caption{$m_{\pi}$-dependence of the real parts of EMR and CMR at
$Q^2=0$ according to non-relativistic 
$\mathcal{O}(\epsilon^3)$ SSE. We note that we have assumed here
that all masses appearing in the definitions of the Jones-Scadron
multipole form factors are taken at their values at the physical
point, in order to display {\em only} the intrinsic quark-mass
dependence of the transition form factors.}
\label{ratiosqm}
\end{center}
\end{figure}

An upcoming task in the description of the nucleon to $\Delta$
transition in chiral effective field theory is the study of the quark-mass dependence of these form factors, extrapolating the recent lattice results of refs.\cite{lattice1},\cite{lattice2} 
to the physical point. Figure \ref{fig:qm}   
shows ---as a first step on this way --- the  pion-mass dependence of the $N\Delta$-transition form factors in the Jones-Scadron basis of Eqs.(\ref{con1}-\ref{con3}) according to non-relativistic
$\mathcal{O}(\epsilon^3)$ SSE. We note that we did not refit any of the parameters of table \ref{fit} to produce the
extrapolation functions shown---all parameters have been fixed from experimental observables at the physical point as
described in sections \ref{fit1}, \ref{fit2}. Both the real and the
imaginary parts of the three $N\Delta$-transition form factors develop
a quark-mass $m_q$-dependence, which has been translated into a dependence on the mass of the pion $m_\pi$ via the GOR-relation \cite{GOR}
\begin{eqnarray}
m_\pi^2 & = & 2\,B_0\,m_q+\mathcal{O}(m_q^2),
\label{GOR}
\end{eqnarray}
consistent with the order at which we are working. $B_0$ denotes the magnitude of the chiral condensate.  The imaginary parts of all three form factors shown in Fig.\ref{fig:qm} vanish\footnote{This vanishing is not
necessarily a monotonous function of $m_\pi$, as can be see in the imaginary part of $\mathcal{G}^*_E(0)$ in Fig.\ref{fig:qm}.} for
$m_{\pi}>\Delta_0$, since the $\Delta$(1232) resonance would become a stable particle at this
large pion-mass. It is interesting to observe, that the quark-mass dependence of the magnetic dipole $N\Delta$-transition moment 
$\mu_{N\Delta}=\textnormal{Re}\left[\mathcal{G}_M^*(0)\right]$ qualitatively
shows the same behaviour as the isovector magnetic moment $\mu_N^v$ of
the nucleon, studied {\it e.g.} in ref.\cite{HW}. Like its analogue $\mu_N^v$, 
at the physical point ($m_\pi=140$ MeV) $\mu_{N\Delta}$ is substantially reduced
from its chiral limit value by $\approx$ 25 percent, dropping\footnote{We assume here that
all masses appearing in Eqs.(\ref{defff}-\ref{con3}) are taken at their physical value, see the discussion in ref.\cite{QCDSFff} regarding this point.}
further rather quickly in size for increasing quark masses. On the other hand, the quark-mass dependence of both the electric and the coulomb
 quadrupole 
$N\Delta$-transition moments
$Q_E^{N\Delta}=\textnormal{Re}\left[\mathcal{G}_E^*(0)\right]$ and 
$Q_C^{N\Delta}=\textnormal{Re}\left[\mathcal{G}_C^*(0)\right]$
is rather unexpected: As can be seen from Fig.\ref{fig:qm} $Q_E^{N\Delta}$ even changes its sign around $m_\pi\approx 100$ MeV, before approaching a {\em negative}
value in the chiral limit. It will be very interesting to see how the
location of this zero-crossing might be affected by corrections at
next-to-leading one-loop order $\mathcal{O}(\epsilon^4)$ in SSE \cite{GH}. In the case of $Q_C^{N\Delta}$ one can observe  the effect of
$G_3(0)$ of Eq.(\ref{chex}), leading to a logarithmic divergence of the coulomb quadrupole transition strength in the chiral
 limit. Curiously, our non-relativistic ${\cal O}(\epsilon^3)$ SSE calculation indicates that $Q_C^{N\Delta}$ is near a local maximum for
physical quark-masses. Given the dominance of chiral $\pi N$-physics in this form factor (see the discussion in section \ref{breakdown}), it would be
extremely exciting if such a behaviour could be observed in a lattice QCD simulation. Unfortunately, present state-of-the-art
lattice simulations for $N\Delta$-transition form factors take place for $m_\pi>370$ MeV \cite{lattice1}, which is outside
the region of applicability\footnote{This ``early'' breakdown of the chiral extrapolation
function for $m_{\pi}>200$ MeV may
come as a surprise, as there are several examples known where non-relativistic ${\cal O}(\epsilon^3)$ SSE calculations have produced
stable extrapolation functions up to a pion mass of 500...600 MeV. (See for example the corresponding SSE calculations for the mass of the 
nucleon in ref.\cite{QCDSFff} or the axial coupling of the nucleon in ref.\cite{HPW}.) For the chiral extrapolation of the $N\Delta$ transition 
form factors, however, it seems that important quark-mass dependent
structures are only generated at ${\cal O}(\epsilon^4)$ in the
non-relativistic SSE calculation,
in particular via pion loop-corrections induced by the coupling $b_1$ \cite{GH}.} of this non-relativistic leading-one-loop SSE calculation,
as indicated by the grey bands in Figs. \ref{fig:qm}, \ref{ratiosqm}. We hope to extend the range in $m_\pi$ of the chiral extrapolation functions
for the $N\Delta$-transition considerably when going to next order in
the calculation \cite{GH}. We also note that the leading-one-loop 
covariant calculation in the $\delta$-expansion scheme presented
in ref.\cite{PV}---which contains the same diagrams as our
non-relativistic ${\cal O}(\epsilon^3)$ SSE calculation, see 
Fig.\ref{fig:feyn}---seems
to be also stable at pion masses larger than $m_\pi\approx 200$ MeV, presumably due to the additional $m_\pi/M_N$ terms present
in a covariant approach. However, the stability of the chiral
extrapolation functions of both schemes should be tested further by going to 
next-to-leading one-loop order.   
 
\begin{figure}
\begin{center}
\includegraphics[width=12cm,clip]{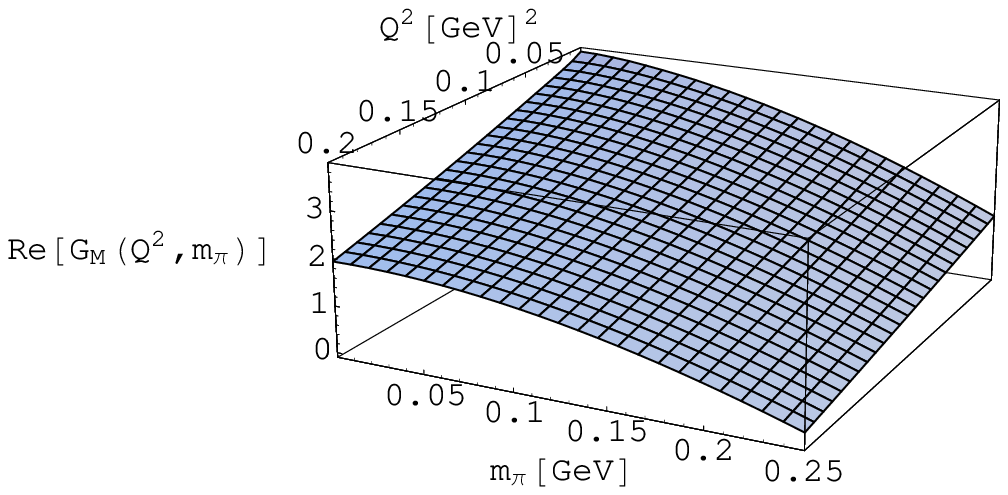}
\caption{$Q^2$ and $m_{\pi}$ dependence of the magnetic dipole $N\Delta$-transition
form factor in non-relativistic ${\cal O}(\epsilon^3)$ SSE.}
\label{GM3D}
\end{center}
\end{figure}

Furthermore, we want to note that there is no need to consider the rather complex structures of EMR or CMR when comparing to lattice
QCD results. The intricacies---{\it i.e.} the sought after signatures of chiral dynamics---can be studied in a much cleaner fashion when directly
comparing ChEFT results to lattice QCD simulations of the $N\Delta$-transition form factors (e.g in the Jones-Scadron basis), see Fig. 
\ref{fig:qm}. Nevertheless, for completeness, in Fig. \ref{ratiosqm} we also show our results\footnote{We note again that all chiral extrapolation functions
shown in this work {\em implicitly} assume that all accompanying mass factors in the definition of the $N\Delta$-transition current
 are held at their physical values. The behaviour of the chiral
extrapolation functions changes significantly for $m_\pi>200$ MeV when the effects of the quark-mass dependence of these masses
are also included \cite{GH}.} for the chiral extrapolation functions of the real parts\footnote{Lattice QCD results
are obtained in Euclidean space and cannot be connected directly with (complex valued) real world observables in Minkowski space in the case
of open decay channels.} of EMR(0) and CMR(0) defined as 
\begin{eqnarray}
emr & = & -\frac{\textnormal{Re}[\mathcal{G}_E^*(0)]}{\textnormal{Re}[\mathcal{G}_M^*(0)]},\\
cmr & = &
	-\frac{M_{\Delta}^2-M_N^2}{4M_{\Delta}^2}
	\frac{\textnormal{Re}[\mathcal{G}_C^*(0)]}{\textnormal{Re}[\mathcal{G}_M^*(0)]} \; .
\end{eqnarray}
The discussed sign-change in $Q_E^{N\Delta}$ is visible in
Fig. \ref{ratiosqm} in {\it emr}, while the exciting chiral structures of $Q_C^{N\Delta}$ 
dominate {\it cmr} for pion masses $m_\pi<200$ MeV. As already mentioned, available lattice data of these quantities in refs.\cite{lattice1,lattice2} are unfortunately
 at too large pion-masses to be relevant for the chiral extrapolation functions presented here. However---independent of this (present) limitation of
 our ChEFT results with respect to small pion-masses---we have to note one important {\em general caveat}
for (future) comparisons of ChEFT results to lattice QCD simulations
of $N\Delta$ form factors: In Figs.\ref{fig:qm},\ref{ratiosqm} we
discuss the chiral extrapolation of the three $N\Delta$-transition
moments at $Q^2=0$, while lattice QCD results
are usually obtained at finite values of momentum transfer. In order
to correct for this, one often attempts to connect the lattice results of finite $Q^2$ to the real photon point by performing global dipole fits, with the dipole mass as a
free (quark-mass dependent) parameter fit from lattice data. While such a procedure may lead to promising results, for example, in the case of
the form factors of the nucleon (see the discussion in ref.\cite{QCDSFff})---and also seems to be applicable to the (monotonously falling) magnetic
dipole $N\Delta$-transition form factor $\mathcal{G}_M^*(Q^2)$--- it should not be applied to the study of the sought after electric and coulomb quadrupole
$N\Delta$-transition form factors, due to the non-trivial momentum
dependence  
in these form factors for $Q^2<0.15$ GeV$^2$. This can be
 clearly concluded from Fig. \ref{fig:main} and from the
negative values of (the real parts of) their radii in table \ref{tab:radii}. Global dipole fits connecting lattice QCD results from large $Q^2$ across
the region $Q^2<0.15$ GeV$^2$ to the photon point at $Q^2=0$ would just "wash-out" all the interesting chiral physics which dominates
 the $N\Delta$ quadrupole transition moments $Q_E^{N\Delta},\,Q_C^{N\Delta}$ at the physical point! If one wants to study these objects
 in lattice QCD one
has to perform simulations at such small values of momentum transfer that one can directly compare with the ChEFT results for the three
$N\Delta$ transition form factors $\mathcal{G}_M^*(Q^2,m_\pi^2)$, $\mathcal{G}_E^*(Q^2,m_\pi^2)$,
$\mathcal{G}_C^*(Q^2,m_\pi^2)$ in the $(Q^2,m_\pi^2)$-plane, and then
utilize a function for the momentum-dependence down to $Q^2=0$
which is consistent with the turning-points generated by chiral $\pi N$-dynamics.
Figure \ref{GM3D} shows the non-relativistic ${\cal O}(\epsilon^3)$
SSE result of such a three dimensional function for $\mathcal{G}_M^*(Q^2,m_\pi^2)$, further quantitative studies
regarding these chiral extrapolation surfaces are relegated to future work \cite{GH}.

\section{Conclusions and Outlook}
\label{sec:conc} 

The pertinent results of our analysis can be summarized as follows:
\begin{enumerate}
\item We have analysed and updated the ${\cal O}(\epsilon^3)$ SSE calculation of the isovector $N\Delta$-transition current of 
ref.\cite{GHKP} in terms of the magnetic dipole, electric quadrupole and coulomb quadrupole transition form factors in Fit 1. It was found
that the momentum-range of reliability of these results is extremely small, $Q^2<0.05$ GeV$^2$.
\item We have identified an ``unnatural'' momentum dependence in the $N\Delta$-transition form factor $G_2(Q^2)$ as the reason for the early
breakdown of the results of Fit 1. In section \ref{fit2} we have demonstrated that the inclusion of a (higher order) counter term, which changes 
the radius of $G_2(Q^2)$ by 0.21 fm, is sufficient to correct the
momentum transfer behaviour in this form factor. 
We have checked that similar correction terms in
the radii of $G_1$ and $G_3$ are not significant. The physical origin
of the short distance contribution to the radius of $G_2$ parametrized in coupling $C_S$ is
not understood at present.
The results of Fit 2 which includes this
correction then showed a consistent behaviour in all three form
factors up to a momentum transfer (squared) of $Q^2=0.25\ldots 0.3$
GeV$^2$.
\item Connecting our results for the transition form factors with ratios of measured pion-electroproduction multipoles via 
Eqs.(\ref{EMR},\ref{CMR}) we have obtained a remarkable agreement between the results of Fit 2 and experiment up to a momentum 
transfer (squared) of $Q^2=0.25\ldots 0.3$ GeV$^2$ both for EMR($Q^2$) and CMR($Q^2$).
\item Long distance pion physics was found to be present in all three $N\Delta$-transition form factors. It showed up most prominent in the 
momentum dependence of the quadrupole form factors for $Q^2<0.15$
GeV$^2$, leading to momentum dependencies which cannot be 
described via a (modified) dipole ansatz anymore. According to our analysis, in pion-electroproduction experiments this feature should show
up most clearly in a rise of CMR for $Q^2<<0.1$ GeV$^2$. It is expected that this rise can be observed within the accuracy of state-of-the art 
pion-electroproduction experiments in the near future \cite{Aron}. Another observable signal of chiral dynamics in the $N\Delta$-transition 
could be a minimum in EMR near $Q^2=0.05$ GeV$^2$ and a maximum near $Q^2=0.25$ GeV$^2$. Unlike the case of CMR, however, 
it is not clear whether these effects can be identified unambiguously given the size of current experimental error bars in EMR$(Q^2)$.
\item We have studied the chiral extrapolation of the three $N\Delta$-transition form factors at $Q^2=0$. We found that the magnetic 
$N\Delta$ dipole transition moment decreases monotonously with the quark-mass, displaying a qualitatively similar behaviour 
as the isovector magnetic moment of the nucleon. On the other hand, the quark-mass dependencies of the 
quadrupole $N\Delta$-transition moments were found to display rapid
changes for pion masses below 200 MeV. While the electric quadrupole
transition moment $Q_E^{N\Delta}$ in our analysis even changes its
sign near $m_\pi=0.1$ GeV before approaching a negative chiral limit value, we found 
that the coulomb quadrupole transition moment $Q_C^{N\Delta}$ has a
local maximum near the physical pion mass, while it diverges in the chiral limit. State-of-the-art
lattice simulations cannot yet reach such small pion masses to test these predictions. On the other hand, our non-relativistic 
${\cal O}(\epsilon^3)$ SSE analysis presented here was found to break
down for $m_\pi>0.2$ GeV.
\item We also want to point out that lattice studies of the
quadrupole $N\Delta$ transition form factors cannot be analyzed via a
simple dipole ansatz to obtain information about the moments at $Q^2=0$ due
to the turning-points and structures in the $Q^2$ dependence of these
form factors at $Q^2<0.1$ GeV$^2$.

\end{enumerate}

In the future we are planning to calculate the 
quark-mass dependence of the $N\Delta$-transition form factors at ${\cal O}(\epsilon^4)$ in order to extend the range of applicability of
these results for chiral extrapolations and to test the stability of
the structure effects discussed in this work \cite{GH}. In conclusion
we can say that we have indeed detected interesting signatures of
chiral dynamics in the $N\Delta$-transition, both for the momentum and
for the quark-mass dependence. In particular we hope that the
electric- and coulomb quadrupole $N\Delta$ transition moments will get tested
with higher precision both on the lattice and in electron scattering
experiments in order to verify the signatures of chiral dynamics
discussed in this work.

\section*{Acknowledgments} 

The authors acknowledge helpful discussions with L. Tiator regarding the experimental data situation. We would also like to thank 
C. Alexandrou and A. Bernstein for valuable input. TAG is  grateful to K. Goeke for helpful financial support in the beginning stages 
of this work.


\begin{appendix}
\section{Integrals}
\label{app:int}

The result of the non-relativistic $\mathcal{O}(\epsilon^3)$ SSE calculation written in
terms of standard loop integrals reads:
\begin{eqnarray}
G_1(q^2) & = &\frac{2c_AM_N}{F_{\pi}^2}\int_0^1\!dx\bigg[g_A
	(x-1)J_2^{\prime}(x\Delta_0,\tilde{m}^2)
	-\frac{5}{3}g_1\left(1-\frac{d-3}{d-1}x\right)J_2^{\prime}(-x\Delta_0,\tilde{m}^2)\bigg] \nonumber
	\\ && -\frac{1}{2}b_1+(2E_1-D_1)\frac{\Delta_0}{4M_N}\\
G_2(q^2) & = &\frac{8c_A M_N^2}{F_{\pi}^2}\int_0^1\!dx\bigg[g_A
	x(x-1)J_1^{\prime}(x\Delta_0,\tilde{m}^2)-
	\frac{5}{3}g_1x(x-1)\frac{d-3}{d-1}J_1^{\prime}(-x\Delta_0,\tilde{m}^2)\bigg]\nonumber\\&&
	+2b_6-D_1\\
G_3(q^2) & = & \frac{4c_AM_N^2\Delta_0}{F_{\pi}^2}\int_0^1\!dx\bigg[g_A
	x(x-1)(1-2x)J_0^{\prime}(x\Delta_0,\tilde{m}^2)-\nonumber \\&&
\frac{5}{3}g_1x(x-1)(2x-1)\frac{d-3}{d-1}J_0^{\prime}(-x\Delta_0,\tilde{m}^2)\bigg].
\end{eqnarray}
The basic loop integrals in d-dimensional regularization are defined as:
\begin{eqnarray}
\frac{1}{i}\int\frac{d^dl}{(2\pi)^d}\frac{1}{m_{\pi}^2-l^2} & = &
\Delta_{\pi} \, = \, 2m_{\pi}^2\left(L+\frac{1}{16\pi^2}\ln{\frac{m_{\pi}}{\lambda}}\right) \\
\frac{1}{i}\int\frac{d^dl}{(2\pi)^d}\frac{1}{(v\cdot l-\omega)(m_{\pi}^2-l^2)}& = &
	J_0(\omega,m_{\pi}^2).
\end{eqnarray}
The analytic expression for the one-nucleon-one-pion loop integral is:
\begin{eqnarray}
J_0(\omega,m_{\pi}^2) & = &
-4L\omega+\frac{\omega}{8\pi^2}\left(1-2\ln{\frac{m_{\pi}}{\lambda}}\right)
	\nonumber \\&&-
	\frac{1}{4\pi^2}\left\{\begin{array}{l c l}
	\sqrt{\omega^2-m_{\pi}^2}\left(\ln{\left(\frac{\omega}{m_{\pi}}+\sqrt{\frac{\omega^2}{m_{\pi}^2}-1}\right)}
	-i\pi\right)&&
	\frac{\omega}{m_{\pi}}>1\\
	\sqrt{m_{\pi}^2-\omega^2}\arccos{\left(-\frac{\omega}{m_{\pi}}\right)} &\,\textnormal{for}\, &
	-1\le\frac{\omega}{m_{\pi}}\le 1 \\
	-\sqrt{\omega^2-m_{\pi}^2}\ln{\left(-\frac{\omega}{m_{\pi}}+\sqrt{\frac{\omega^2}{m_{\pi}^2}-1}\right)}&&
	\frac{\omega}{m_{\pi}}<-1\end{array}\right.
\end{eqnarray}

\begin{eqnarray}
J_0^{\prime}(\omega,m_{\pi}^2) & = & -\frac{\partial}{\partial m_{\pi}^2}J_0(\omega,m_{\pi}^2),\\
J_1^{\prime}(\omega,m_{\pi}^2) & = &
	-\frac{\partial}{\partial m_{\pi}^2}\left(\omega J_0(\omega,m_{\pi})+\Delta_{\pi}\right),\\ 	
J_2^{\prime}(\omega,m_{\pi}^2) & = &
	-\frac{1}{d-1}\frac{\partial}{\partial m_{\pi}^2}
	\left[(m_{\pi}^2-\omega^2)J_0(\omega,m_{\pi})-\omega \Delta_{\pi}\right].
\end{eqnarray}
The divergences at $d=4$ parametrized via dimensional regularization
are collected in the function $L$ in the $\overline{\textnormal{MS}}$-scheme:
\begin{eqnarray}
L & = & \frac{\lambda^{d-4}}{16\pi^2}\left[\frac{1}{d-4}+\frac{1}{2}\left(\gamma_E-1-\ln{4\pi}\right)\right]. 
\end{eqnarray}

\section{The coupling constant $c_A$} \label{app:ca}

We determine the strength of the $N\Delta$ axial coupling constant $c_A$ from the
strong decay width of the $\Delta(1232)$ resonance at tree level. In the rest frame of the $\Delta$ this width
reads:
\begin{eqnarray}
\Gamma_{\Delta\rightarrow N\pi} & = & \frac{c_A^2}{6\pi
F_{\pi}^2}\left(E_{\pi}^2-m_{\pi}^2\right)^{\frac{3}{2}}\cdot
\frac{M_{\Delta}+M_N-E_{\pi}}{2M_{\Delta}},
\end{eqnarray}
where the $\pi N\Delta$-vertex has been taken from Eq.(\ref{Lag4}) and the
associated pion energy function reads:
\begin{eqnarray}
E_{\pi} & = & \frac{M_{\Delta}^2-M_N^2+m_{\pi}^2}{2M_{\Delta}}.
\end{eqnarray}
For the numerical determination of $c_A$ we use the parameters $M_{\Delta}$, $M_N$, $m_{\pi}$ and $F_{\pi}$
from table \ref{tab:param}, a width of
$\Gamma_{\Delta\rightarrow N\pi}=100$ MeV \cite{PDG} and arrive at the result $c_A=1.5$.

\end{appendix}


\end{document}